\documentclass[aps,prd,preprint,groupedaddress]{revtex4-1}

\usepackage{amsmath,amssymb,graphics,epsfig,subfigure}
\usepackage{color}

\begin{document}


\title{ Gravity and Matters on a pure geometric thick polynomial $f(R)$ brane }


\author{Heng Guo \footnote{Corresponding author}}
\email[]{hguo@xidian.edu.cn}
\affiliation{School of Physics, Xidian University, Xi'an 710071, China}
\author{Lang-Lang Wang }
\email[]{wllsrr@163.com}
\affiliation{School of Physics, Xidian University, Xi'an 710071, China}
\affiliation{School of General Education, Shanxi Institute of Science and Technology,
 Jincheng 048000, China}
\author{Chun-E Fu }
\email[]{fuche13@mail.xjtu.edu.cn}
\affiliation{School of Science, Xi'an Jiaotong University, Xi'an 710049, China}
\author{Qun-Ying Xie }
\email[]{xieqy@lzu.edu.cn}
\affiliation{School of Information Science and Engineering, Lanzhou University, Lanzhou 730000, China}


\date{\today}

\begin{abstract}
 In this paper, writing the most general form $f(R)=\sum_{i=1}^{n}a_{i}R^{i}+\Lambda$ the solutions of the pure geometric thick $f(R)$ brane are investigated. For the certain value of $n$, the analytical thick brane solution can be calculated, and when $n=3$, $n=4$, and $n=10$, the thick brane solutions are presented. The solutions are stable against linear tensor perturbations. The zero mode of gravity and scalar field can be localized on thick $f(R)$ branes naturally. The zero mode of vector field and left-chiral fermion can be localized on thick $f(R)$ branes by introducing the coupling with scalar curvature $R$ of spacetime, and the massive resonant modes can be quasilocalized on the brane with the large coupling coefficients.

\end{abstract}


\maketitle

\section{\label{sec:intro} Introduction}

 The idea that our four-dimensional (4D) Universe can be considered as a brane embedded in a higher-dimensional spacetime, can supply new insights for solving the gauge hierarchy problem \cite{NSG1998,INSG1998,LR1999A,M2002,JGM2010,JM2011,KYYXS2012,HAYDR2013} and the cosmological constant problem \cite{VM1983,VM1983E,SC1986,E1986,CJCT2000,NSNR2000,SME2000,BG2000,JBH2001,GDM2001,AK2002,A2004,I2011}. In the Randall-Sundrum (RS) braneworld model \cite{LR1999}, the effective 4D gravity could be recovered even in the case of noncompact extra dimensions, however, singularities are present at the position of the branes. The smooth thick braneworld solutions are generally based on gravity coupled to bulk scalar fields \cite{FCR1999,WM1999,M2000,ODSA2000,CJTY2000,S2001,S2001A,A2002,RA2002,SKJ2002,KB2003,MN2003,ANA2003,DCA2004,
DA2004,DFL2006,VDL2006,V2007,VDRA2007,DR2007,VVDS2008,VVM2009,DALR2009,DAL2009,AM2010,ZYHY2010,YYK2010,YYZH2011,
HYZH2011,HYZF2012,HYSC2012,DFF2013,AGJ2014,GADIR2014,DARAA2014,DAR2015,DALRG2015,DLRGD2015,BYK2015,LVNB2017}.
For some comprehensive reviews about thick branes, please see Refs. \cite{S2008,VVM2010,M2010,THLRH2010,RK2010}. There are also thick branes arising from pure geometry without the inclusion of bulk scalar fislds at all \cite{ORI2002,NA2005,NA2006,NAMC2008,YKY2010,VVBJ2010,HHZ2012,JYYY2012,NAKUI2014,YY2016}.

In braneworld scenarios, the important issue is the localization of gravity and various bulk matter fields for the purpose of recovering the effective 4D gravity and building up the standard model. Generally the gravity and the scalar field zero mode can be localized on the brane naturally. For five-dimensional (5D) free spin-1 Abelian vector field, the zero mode can not be trapped on the Minkowski ($\mathcal{M}_{4}$) brane , however, it can be localized on the thick de Sitter brane and Weyl thick brane \cite{YXLY2008,YLLY2008,YLSY2008,YHCH2011,HAYDR2013}. In some ways, by introducing the coupling with background geometry, the zero mode of the vector field can also be trapped on the Minkowski brane \cite{ZQY2015,ZQ2018}. For the spin-1/2 fermion, zero mode cannot be localized on the branes without introducing the coupling. By introducing the coupling between the fermion and background scalar fields or fermion and background geometry, the left-chiral fermion zero mode can be localized on the brane \cite{BG2000,SM2000,SVP2000,TM2000,YM2000,AK2001,I2000,I2000C,I2002,S2002,CPJ2002,AVPR2003,AVPR2005,RS2005,RS2005S,YTYY2005,ANJ2006,AM2007,TR2007,YLY2007,YLXY2007,YXLY2008,YLLY2008,YT2008,LYY2008,YLSY2008,MWC2009,AM2009F,YKN2009,YJZCY2009,YCLY2009,YZSY2009,CRMA2009,RJS2009,ZYHY2010,L2011,AAM2011,WAC2011,CYH2011,OI2012,TN2012,HAYDR2013,AVO2013,V2013,MP2014,LCDC2014,YZFS2014,NDR2015,HQC2015,IGRR2015,RAGAR2016,YYWY2017}.

However, due to the fact that Einstein's general relativity is not renormalizable, the effects of higher order curvature terms are suggested to consider. $f(R)$ gravity theories, which the Lagrangians are proportional to some functions of the scalar curvature $R$, were created in the investigation of cosmology, and offer pure geometric explanations for cosmologial inflation and dark energy in physically \cite{VLD1987,JS1988,GESSLS2008,AS2010,TV2010,SS2011,NDepjc2018,NDplb2018}.
Nevertheless, there are studies devoted to embedding branes into various types of $f(R)$ gravities \cite{MSD2005,VDRA2007,NMY2008,AM2008,AM2009,VVBJ2010,JM2011,YYZH2011,YYK2011,HHZ2012,
TMJ2013,DRAA2013,DARAA2014,DLRGD2015,DAR2015,DALRG2015,ZYHY2015,BBHY2015,HYBY2016,RMLS2015,VDVF2019,YY2016,CuiJHEP}. In Ref. \cite{YY2016}, two cases of the thick brane solutions have been investigated in pure geometry $f(R)$ gravities: one with a triangular $f(R)$ and the other a simple polynomial $f(R)$.

In this paper, we intend to investigate the thick branes generated by only geometry under pure $f(R)$ gravity. The general form of $f(R)$ would be considered, which is chosen as a $n$-th order polynomial of the scalar curvature $R$. When the degree of the polynomial $f(R)$ is $n = 3$, the solution is equivalent to the one of Ref. \cite{YY2016}, When $n = 4$ and $n = 10$, the solutions are new solutions. These solutions will be presented in next section. The stability of tensor perturbations and the localization of gravity are discussed in Sec.\ref{sec:gravity}. The zero mode of gravity can be localized on the brane naturally. In Sec.\ref{sec:scalar}, the zero mode of the scalar field can be trapped on the branes naturally. In Sec.\ref{sec:vector}, by introducing the coupling between the vector field and the background geometry, the zero mode vector can be localized on the brane and the massive resonant Kaluza-Klein (KK) modes can also be quasilocalized on the brane. In Sec.\ref{sec:fermion}, for a pure geometry brane, the coupling between the fermion and the background scalars can not be introduced, however, we introduce the coupling between the fermion and the background geometry, and the zero mode of the left-chiral fermion and the massive resonant KK modes can be localized and  quasilocalized on the brane. Finally, the conclusion and discussion are given in Sec.\ref{sec:conclusion}.

\section{\label{sec:fR branes} The pure geometric thick $f(R)$ branes}

We star with the following 5D action for a pure geometric thick $f(R)$ braneworld
\begin{eqnarray}\label{gravity action}
    S = \frac{1}{2\kappa_{5}^{2}} \int d^{5}x \sqrt{-g} f(R),
\end{eqnarray}
where $\kappa_{5}^{2} = 8\pi G_{5}$ with $G_{5}$ being the 5D Newton constant, $R$ is the 5D scalar curvature, and $g = \det(g_{MN})$ is the determinant of the metric. Throughout this paper, capital Latin letters $M, N, \cdots = 0,1,2,3,5$ and Greek letters $\mu,\nu,\cdots = 0,1,2,3 $ are used to represent the bulk and brane indices, respectively. From this action, the 5D Einstein equation is given by
\begin{eqnarray}
    R_{MN}f_{R} -\frac{1}{2}g_{MN}f(R) +(g_{MN}\Box -\nabla_{M}\nabla_{N})f_{R} = 0, \label{Einstein Eq}
\end{eqnarray}
where $f_{R}\equiv \frac{df(R)}{dR}$, $\Box = g^{MN}\nabla_{M}\nabla_{N}$ is the 5D d'Alembert operator, and $R_{MN}$ is the 5D Ricci tensor, defined in terms of the Riemann tensor $R_{MN} = R^{Q}_{MQN}$.

The line element of the $\mathcal{M}_{4}$ brane is assumed as
\begin{eqnarray}
    ds^{2} = g_{MN}dx^{M}dx^{N} = \text{e}^{2A(y)} \eta_{\mu\nu} dx^{\mu}dx^{\nu} + dy^{2}, \label{metric}
\end{eqnarray}
where $\text{e}^{2A(y)}$ is the warp factor,
\begin{eqnarray}
 \eta_{\mu\nu} = \text{diag} (- 1, + 1, + 1, + 1)
\end{eqnarray}
is the metric of the 4D Minkowski spacetime, and $y$ denotes the extra dimensional coordinate. From the metric (\ref{metric}), the Ricci tensor and scalar curvature can be computed:
\begin{eqnarray}
    R_{\mu\nu} &=& -\text{e}^{2A(y)}(A''+4A'^{\;2})\eta_{\mu\nu}, \quad\quad
    R_{55} = -4A''-4A'^{\;2}, \label{Ricci tensor} \\
    R &=& -20A'^{\;2}-8A'', \label{Ry}
\end{eqnarray}
where the prime denotes the derivatives with respect to extra dimensional coordinate $y$. The Einstein equations (\ref{Einstein Eq}) can be rewritten as:
\begin{subequations}\label{Eeq}
\begin{eqnarray}
    f(R)+2f_{R}(4A'\,^{2}+A'')-6f^{\;'}_{R}A'-2f^{\;''}_{R} = 0, \label{Einstein Eq1}\\
    -f(R)-8f_{R}(A''+A'^{\;2})+8f_RA' = 0. \label{Einstein Eq2}
\end{eqnarray}
\end{subequations}
Above two equations are not independent, because the left side of Eq. (\ref{Einstein Eq}) is divergence free \cite{TV2010}. Hence, we choose to solve Eq. (\ref{Einstein Eq1}) in the following discussion.

In this paper, we are interested in investigating the general form of $f(R)$, chosen as a $n$-th order polynomial of the scalar curvature $R$:
\begin{eqnarray}
    f(R) = \sum_{i = 1}^{n} a_{i}\;R^{\,i} +\Lambda, \label{fR}
\end{eqnarray}
where the $a_{i}$ coefficients have the appropriate dimensions, $\Lambda$ is 5D cosmological constants, and the parameter $n$ should be an integer.

\begin{figure}
 \centering
\includegraphics[width = 0.4\textwidth]{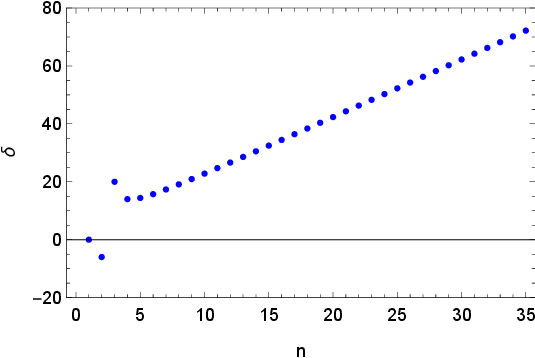}
\caption{The relation between the parameter $\delta$ and $n$.}
    \label{fig_delta}
\end{figure}

According to Eqs. (\ref{Ry}) and (\ref{fR}), one can easily prove that $f(R)$ includes $(A'')^{n}$ and $(A')^{2n}$, $f^{\;'}_{R}$ includes $A'''$ and $f^{\;''}_{R}$ includes $A''''$. By taking $f(R)$, $f^{\;'}_{R}$ and $f^{\;''}_{R}$ back into Eqs. (\ref{Eeq}), obviously, the Einstein equations are high-order differential equations, and very hard to be solve.
In order to find a solution, we begin with a simple $A(y)$, which consider as \cite{YY2016,CuiJHEP}
\begin{eqnarray}
    A(y) = -\delta \ln(\cosh(ky)) \label{Ay}
\end{eqnarray}
with $\delta$ a dimensionless positive constant, and $k$ another positive constant with the dimension of length inverse.
 At the infinity, the behavior of the $A(y\rightarrow \infty)$ is
\begin{eqnarray}\label{Ay behavior}
   A(y\rightarrow \infty) \rightarrow -\delta k |y| .
\end{eqnarray}
Defining a new parameter by $\tilde{k}=\delta k$, it is easy to proof that the warp factor
\begin{eqnarray}
 \text{e}^{2A(y\rightarrow \infty)}\rightarrow\text{e}^{-2\tilde{k} |y|},
\end{eqnarray}
 which is the same as the warp fact of the RS II braneworld and implies that the metric (\ref{metric}) reduces an Anti-de Sitter (AdS) one, which is essential for the localization of gravitation.
 The following expressions can also be taken
\begin{eqnarray}
    A'(y) &=& -\delta k\tanh(ky), \\
    A'^{2}(y) &=& \delta^{2}k^{2}(1-\text{sech}^{2}(ky)), \\
    A''(y) &=& -\delta k^{2}\text{sech}^{2}(ky),
\end{eqnarray}
and the scalar curvature takes a simple form:
\begin{eqnarray}
    R(y) = 20\delta^{2}k^{2} (\varepsilon\;\text{sech}^{2}(ky)-1), \label{RyEq}
\end{eqnarray}
where $\varepsilon = \frac{5\delta+2}{5\delta}$.
Furthermore, we can obtain the following expressions
\begin{eqnarray}
    f(R)
    =
    \sum_{i = 1}^{n} a_{i} (20\delta^{2}k^{2})^{i} (\varepsilon\,\text{sech}^{2}(ky)-1)^{i} + \Lambda,
\end{eqnarray}
\begin{eqnarray}
    f_{R}
    =
    \sum^{n}_{i = 1} i \,a_{i} (20\delta^{2}k^{2})^{i-1}
    (\varepsilon \,\text{sech}^{2}(ky) - 1)^{i-1},
\end{eqnarray}
\begin{eqnarray}
    f^{\;'}_{R}
    &=&
    \sum^{n}_{i = 1} 2i(1-i) a_{i} k \varepsilon \,(20\delta^{2}k^{2})^{i-1} \text{sech}^{2}(ky) \tanh(ky) \nonumber \\
    &&~
    \times(\varepsilon \,\text{sech}^{2}(ky) - 1)^{i-2},
\end{eqnarray}
\begin{eqnarray}
    f^{\;'}_{R} A'
    &=&
    \sum^{n}_{i = 1} 2i(i-1) a_{i} \delta k^{2i} \varepsilon \,(20\delta^{2})^{i-1} \text{sech}^{2}(ky) \nonumber \\
    &&~
    \times(1 - \text{sech}^{2}(ky)) (\varepsilon\,\text{sech}^{2}(ky) - 1)^{i-2} ,
\end{eqnarray}
\begin{eqnarray}
    f^{\;''}_{R}
    &&=
    \sum^{n}_{i = 1} 2i(i - 1)a_{i}\varepsilon k^{2i}(20\delta^{2})^{i - 1}
     (\varepsilon\,\text{sech}^{2}(ky)-1)^{i-3}    \nonumber \\
    &&~
    \times \text{sech}^{2}(ky)[ 2(i-2)\varepsilon\text{sech}^{2}(ky)(1-\text{sech}^{2}(ky)) \nonumber \\
    &&~~~~~
    +(2-3\text{sech}^{2}(ky))
    (\varepsilon\,\text{sech}^{2}(ky)-1)]
\end{eqnarray}
By taking above expressions back into Eq. (\ref{Einstein Eq1}), it is clear that all the terms of the Einstein equation are the function of $\text{sech}^{2}(ky)$, and the order of the highest is $2n$. The Einstein equation (\ref{Einstein Eq1}) can be expressed as the following algebraic equation:
\begin{eqnarray}
    \sum_{i = 0}^{n} B_{i}\;\text{sech}^{2i}(ky) = 0, \label{B_sech}
\end{eqnarray}
where the coefficient $B_{i}$ can be written as follows
\begin{eqnarray}
    B_{i}
    &=&
    \sum_{j = i}^{n} (-1)^{j-i} 2a_{j} (20\delta^{2})^{j-1} k^{2j} \varepsilon^{i-1} \nonumber \\
    &&~
    \times\Big[ 10\delta^{2} \text{C}_{j}^{\,i} \,\varepsilon
    - j \,\delta \Big(  4\delta \text{C}_{j-1}^{\,i} \,\varepsilon
    + ( 1 + 4\delta) \text{C}_{j-1}^{\,i-1} \Big) \nonumber \\
    &&~
    + 2 j (j-1) \Big( (2 + 3\delta) \text{C}_{j-2}^{\,i-1} \,\varepsilon + (3 + 3\delta) \text{C}_{j-2}^{\,i-2} \Big) \nonumber \\
    &&~
    + 4j (j-1) (j-2) (\text{C}_{j-3}^{\,i-2} \,\varepsilon + \text{C}_{j-3}^{\,i-3}) \Big],    \label{Bi}
\end{eqnarray}
with $a_{0} \equiv \Lambda$ and $\text{C}^{\,i}_{j}$ the binomial coefficient.
Here we define $\text{C}^{\,i}_{j} \equiv 0$ when $i > j\,$.
 The $B_{i}$ can also be expressed as a column matrix:
\begin{eqnarray}
    \left(
\begin{matrix}
    B_{0}\\
    B_{1}\\
    \vdots\\
    B_{n-1}\\
    B_{n}
\end{matrix}
    \right)
    =
    \left(
\begin{matrix}
    b_{0,0}&   b_{0,1}&   \dots&   b_{0,n-1}&   b_{0,n}\\
    b_{1,0}&   b_{1,1}&   \dots&   b_{1,n-1}&   b_{1,n}\\
    \vdots&    \vdots&    b_{i,j}& \vdots&      \vdots\\
    b_{n-1,0}& b_{n-1,1}& \dots&   b_{n-1,n-1}& b_{n-1,n}\\
    b_{n,0}&   b_{n,1}&   \dots&   b_{n,n-1}&   b_{n,n}
\end{matrix}
    \right)_{((n+1)\times(n+1))} \times
    \left(
\begin{matrix}
    a_{0}\\
    a_{1}\\
    \vdots\\
    a_{n-1}\\
    a_{n}
\end{matrix}
    \right), \label{ba}
\end{eqnarray}
where the element of the matrix $b_{i,j}$ is
\begin{eqnarray}
    b_{i,j}
    &=&
    (-1)^{j-i} 2(20\delta^{2})^{j-1} k^{2j} \varepsilon^{i-1} \nonumber \\
    &&~
    \times\Big[ 10\delta^{2} \text{C}_{j}^{\,i} \,\varepsilon
    - j \,\delta \Big( 4\delta \text{C}_{j-1}^{\,i} \,\varepsilon
    + ( 1 + 4\delta) \text{C}_{j-1}^{\,i-1} \Big) \nonumber \\
    &&~
    + 2 j (j-1) \Big( (2 + 3\delta) \text{C}_{j-2}^{\,i-1} \,\varepsilon + (3 + 3\delta) \text{C}_{j-2}^{\,i-2} \Big) \nonumber \\
    &&~
    + 4j (j-1) (j-2) (\text{C}_{j-3}^{\,i-2} \,\varepsilon + \text{C}_{j-3}^{\,i-3}) \Big].
    \label{bij}
\end{eqnarray}
By setting all the coefficients to zero $B_{i} = 0$, the solution to the Eq (\ref{B_sech}) can be calculated.
Analyzing the expression (\ref{bij}), when $i>j$, it is easy to find that $b_{i,j} = 0$, since all the binomial coefficients are zero. Specially, except for $b_{n,n}$, the elements of the last row of the matrix are zero, so the coefficient $B_{n}$ satisfies the following formulation
\begin{eqnarray}
    B_{n}
    &=&
    2k^{2} a_{n}(20\delta^{2}k^{2})^{n-1} \varepsilon^{n-1} \Big[ (10 - 4n)\delta^{2} \nonumber \\
    &&
    + (6n^{2} - 7n + 4) \delta + (4n^{3} - 6n^{2} + 2n)\Big] =0   \label{delta Eq}.
\end{eqnarray}
For non-zero solution of $a_{n}$, the relationship between $\delta$ and $n$ can be calculated as follows
\begin{eqnarray}
    \delta = \frac{4n^{2}-6n+2}{2n-5}. \label{delta}
\end{eqnarray}
For each certain value of $n$, the parameter $\delta$ can be obtained as it is depicted in Fig.\,\ref{fig_delta}. In order to ensure that the behavior of warp factor at infinity tends to zero, $\delta$ must be positive definite, so the minimum value of the positive integer $n$ is $n_{\text{min}} = 3$. It is easy to see that when $n = 4$, the minimun $\delta$ can be achieved $\delta_{\text{min}} = 14$, and when $n$ tends to infinity, $\delta$ is proportional to $n$ and tends to divergence.

Furthermore, Eq. (\ref{ba}) can be reexpressed as follows
\begin{equation}\label{Bnbij}
\left(
\begin{matrix}
    1&      b_{0,1}&   \dots&   b_{0,n-1}&   b_{0,n}\\
    0&      b_{1,1}&   \dots&   b_{1,n-1}&   b_{1,n}\\
    \vdots& \vdots&    b_{i,j}& \vdots&      \vdots\\
    0&        0&       \dots&   b_{n-1,n-1}& b_{n-1,n}\\
    0&        0&       \dots&   0&           0
\end{matrix}
    \right)_{(n+1)\times(n+1)} \times
    \left(
\begin{matrix}
    \Lambda\\
    a_{1}\\
    \vdots\\
    a_{n-1}\\
    a_{n}
\end{matrix}
    \right)
    =
    \left(
\begin{matrix}
    0\\
    0\\
    \vdots\\
    0\\
    0
\end{matrix}
    \right).
\end{equation}
By considering the Eqs. (\ref{bij}) and (\ref{delta}), all the elements of the matrix $b_{i,j}$ can be calculated for the given value of $n$, so the matrix equation (\ref{Bnbij}) contains $n$ linear equations and $n + 1$ unknown parameters $\Lambda, a_{1}, a_{2}, a_{3}, \cdots, a_{n}$, and the relationships of these parameters could be concluded by solving Eq. (\ref{Bnbij}).

When $n = 1$ and $f(R) = a_{1}R +\Lambda$, the case of general relativity is recovered. Because $\delta = 0$ and $A(y) = 0$, there is no pure geometric thick $\mathcal{M}_{4}$ brane solution. However, by using a different metric ansatz, a de Sitter $d S_{4}$ brane solution can be obtained, which has been studied in Refs. \cite{HAYDR2013,ADR2010}, or by introducing the tension of the brane, the RS braneworld model \cite{LR1999,LR1999A} can also be concluded.

When $n = 2$ and $f(R) = a_{1}R +a_{2}R^{2} +\Lambda$, the parameter $\delta = -6$ and $A(y) = 6\ln(\cosh(ky))$, thus the warp factor $\text{e}^{2A(y)} = \cosh^{12}(ky)$ is divergent at infinity of the extra dimensional coordinate. By introducing a background scalar, the divergence of warp factor is removed and the thick branes can be constructed in Ref. \cite{YYZH2011,DARAA2014,ZYHY2015}.

\begin{figure}
\begin{center}
\includegraphics[width = 0.5\textwidth]{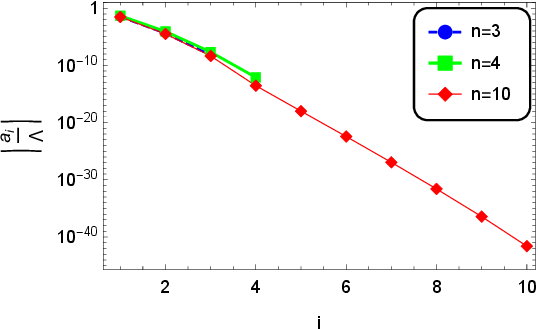}
\caption{The coefficients $|\frac{a_{i}}{\Lambda}|$ in logarithmic coordinate. $n = 3$ for the dashed blue line, $n = 4$ for the green line, and $n = 10$ for the thin red line. The parameter is set to $k = 1$.}
   \label{fig_ai}
\end{center}
\end{figure}

When $n=3$, the parameter $\delta = 20$ and warp factor $A(y)=-20\ln(\cosh(ky))$, and a pure geometric thick $\mathcal{M}_{4}$ brane can be received. When $n=4$, the parameter $\delta$ can be achieved the minimum value $\delta=\delta_{\text{min}}=14$, and a pure geometric thick $\mathcal{M}_{4}$ brane can also be constructed.
When $n=5, 6, \cdots$, the solutions of a thick $\mathcal{M}_{4}$ brane can be concluded, however, as the value of the parameter $n$ increases, the calculation also increases. In this paper, the solutions will be discussed when $n = 3$ (for the minimum $n$), $n = 4$ (for the minimum $\delta$), and $n = 10$ (for a large value $n$), and these cases are very typical. The parameters of the solutions are shown in Table \ref{table ai}. Here, the coefficient $a_{1}$ is set to positive, so the 5D cosmological constant must be negative $\Lambda < 0$, and the 5D spacetime is AdS one. Moreover, the coefficients $a_{i}$ for the solutions are shown in Fig.\,\ref{fig_ai}, and it can be found that the coefficients $a_{i}$ significantly decreases with increasing the order $i$. For the case of $n = 3$, if we set $\Lambda = -\frac{377600k^{2}}{7803}$, the result is the same as example 2 in Ref.\cite{YY2016}.

\begin{table}[tbp]
\centering
\begin{tabular}{|c|c|c|c|}
\hline
    $n$      & $n = 3$       & $n = 4$       & $n = 10$  \\
\hline
    $\delta$ & $\delta = 20$ & $\delta = 14$ & $\delta = \frac{114}{5}$  \\
\hline
    $a_{1}$  &
    $- \frac{3147}{94400k^{2}}\Lambda \approx -0.0333 \frac{\Lambda}{k^{2}}$ &
    $- 0.0478 \frac{\Lambda}{k^{2}}$ &
    $- 0.0292 \frac{\Lambda}{k^{2}}$
    \\
    $a_{2}$  &
    $\frac{249}{6041600k^{4}}\Lambda \approx 4.121 \times 10^{-5} \frac{\Lambda}{k^{4}}$ &
    $2.488 \times 10^{-5} \frac{\Lambda}{k^{4}}$ &
    $3.166 \times 10^{-5} \frac{\Lambda}{k^{4}}$
    \\
    $a_{3}$  &
    $- \frac{13}{1933312000k^{6}}\Lambda \approx -6.724 \times 10^{-9} \frac{\Lambda}{k^{6}}$ &
    $- 2.166 \times 10^{-8} \frac{\Lambda}{k^{6}}$ &
    $- 4.403 \times 10^{-9} \frac{\Lambda}{k^{6}}$
    \\
    $a_{4}$  &               &
    $7.927 \times 10^{-13} \frac{\Lambda}{k^{8}}$ &
    $- 2.834 \times 10^{-14} \frac{\Lambda}{k^{8}}$
    \\
    $a_{5}$  &               &               &
    $- 1.002 \times 10^{-18} \frac{\Lambda}{k^{10}}$
    \\
    $a_{6}$  &               &               &
    $- 3.689 \times 10^{-23} \frac{\Lambda}{k^{12}}$
    \\
    $a_{7}$  &               &               &
    $- 1.130 \times 10^{-27} \frac{\Lambda}{k^{14}}$
    \\
    $a_{8}$  &               &               &
    $- 2.550 \times 10^{-32} \frac{\Lambda}{k^{16}}$
    \\
    $a_{9}$  &               &               &
    $- 3.684 \times 10^{-37} \frac{\Lambda}{k^{18}}$
    \\
    $a_{10}$ &               &               &
    $- 2.530 \times 10^{-42} \frac{\Lambda}{k^{20}}$ \\
\hline
\end{tabular}
\caption{\label{table ai} The solutions of the pure geometric $f(R)$ thick brane for $n=3$, $n=4$, and $n=10$.}
\end{table}

\begin{figure*}
\begin{center}
\subfigure[]{\label{fig_Ay}
\includegraphics[width=0.4\textwidth]{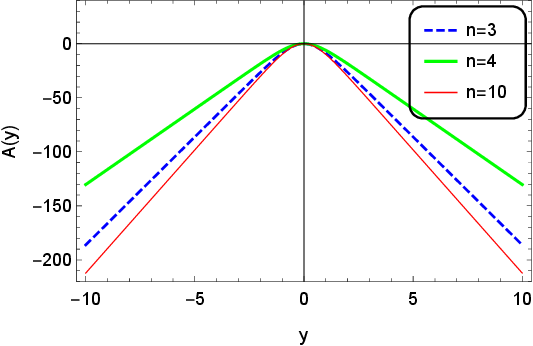}}
\subfigure[]{\label{fig_Ry}
\includegraphics[width=0.4\textwidth]{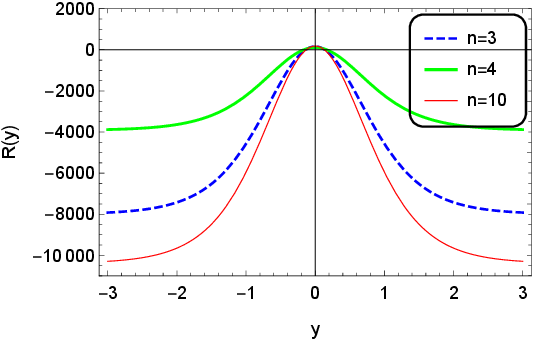}}
\end{center}\vskip -5mm
\caption{The shapes of $A(y)$ in (a) and $R(y)$ in (b) for different thick brane solutions.
$n = 3$ for the dashed blue line, $n = 4$ for the green line, and $n = 10$ for the thin red line. The parameter is set to $k = 1$.}
\label{fig_AyRy}       
\end{figure*}

The pure geometric thick branes can be constructed, and the behaviors of the scalar curvature (\ref{RyEq}) at zero and infinity can be analyzed as follows:
\begin{eqnarray}\label{Ry behavior}
\begin{cases}
   R(y = 0) = 8\delta k^{2} ,  \\
   R(y \rightarrow \infty) \rightarrow -20\delta^{2}k^{2}.
\end{cases}
\end{eqnarray}
The shapes of $A(y)$ and $R(y)$ are shown in Figs.\,\ref{fig_Ay} and \ref{fig_Ry}. We can find that both of $A(y)$ and $R(y)$ are smooth of the extra dimensional and $R(y)$ tends to a constant $-20\delta^{2}k^{2}$ at infinity.

\section{\label{sec:gravity} Tensor perturbation and the localization of gravity on the thick $f(R)$ branes}

The stability of the tensor perturbations of the gravity will be studied in this section. The small perturbations $h_{\mu\nu}$ are introduced into the metric \cite{YYK2011}:
\begin{eqnarray}
    ds^{2} = \text{e}^{2A(y)}(\eta_{\mu\nu} + h_{\mu\nu})dx^{\mu}dx^{\nu} + dy^{2}
\end{eqnarray}
where $h_{\mu\nu} = h_{\mu\nu}(x^{\rho},y)$ depend on all the spacetime coordinates. The following relations can be obtained immediately:
\begin{eqnarray}
    \delta R_{\mu\nu}
    &=&
    - \frac{1}{2} (\hat{\Box}h_{\mu\nu} + \partial_{\mu}\partial_{\nu}h
    - \partial_{\nu}\partial_{\sigma}h^{\sigma}_{\mu} - \partial_{\mu}\partial_{\sigma}h^{\sigma}_{\nu}) \nonumber \\
    &&~
    - 2\text{e}^{2A}A'h'_{\mu\nu} - 3h_{\mu \nu}\text{e}^{2A}A^{'\,2} -\frac{1}{2}\text{e}^{2A}h''_{\mu\nu} \nonumber \\
    &&~
    - \frac{1}{2}\text{e}^{2A}A'\eta_{\mu\nu}h',
    \label{delta R1} \\
    \delta R_{\mu 5}
    &=&
    \frac{1}{2}\partial_{y}(\partial_{\lambda}h^{\lambda}_{\mu}-\partial_{\mu}h), \label{delta R2}\\
    \delta R_{5 5}
    &=&
    - \frac{1}{2}(2A'h'+h''),
    \label{delta R3}\\
    \delta R
    &=&
    \delta(g^{\mu\nu}R_{\mu\nu}) \nonumber \\
    &=&
    -\text{e}^{-2A}\hat{\Box}h + \text{e}^{-2A}\partial_{\mu}\partial_{\nu}h^{\mu \nu} - 5A'h' - h'', \label{delta R4}
\end{eqnarray}
Where $\hat{\Box} = \eta^{\mu\nu}\partial_{\mu}\partial_{\nu}$ is the 4D d'Alembert operator, and
$h = \eta^{\mu\nu}h_{\mu\nu}$. The tensor perturbations satisfy the transverse-traceless (TT) condition $\partial_{\mu}h^{\mu}_{\nu} = h = 0 $.

The perturbed Einstein equations can be expressed:
\begin{eqnarray}
    &&
    \delta R_{MN}f_{R} + R_{MN}f_{RR}\delta R - \frac{1}{2}\delta g_{MN}f(R)  \nonumber \\
    &&
    - \frac{1}{2}g_{MN}f_{R}\delta R + \delta(g_{MN}\Box f_{R}) - \delta(\nabla_{M}\nabla_{N}f_{R}) = 0, \label{P E}
\end{eqnarray}
where
\begin{eqnarray}
    f_{RR} \equiv \frac{\text{d}f_{R}}{\text{d}R} = \frac{\text{d}^{2}f(R)}{\text{d}R^{2}}, \nonumber
\end{eqnarray}
\begin{eqnarray}
    \delta(\nabla_{M}\nabla_{N}f_{R})
    &=&
    (\partial_{M}\partial_{N} -\Gamma^{P}_{MN}\partial_{P}) (f_{RR}\delta R)
    - \delta\Gamma^{P}_{MN}\partial_{P}f_{R},  \\
    \delta(g_{MN}\Box f_{R})
    &=&
    \delta g_{MN}\Box f_{R} + g_{MN}\delta g^{AB} (\nabla_{A}\nabla_{B}f_{R})
    + g_{MN}g^{AB} \delta(\nabla_{A}\nabla_{B}f_{R}).
\end{eqnarray}
By substituting the Eqs. (\ref{delta R1}),(\ref{delta R2}),(\ref{delta R3}), and (\ref{delta R4}) into the Eq.(\ref{P E}), the perturbed Einstein equations can be simplified:
\begin{eqnarray}
    \Big( \text{e}^{-2A}\hat{\Box}h_{\mu\nu} + 4A'h'_{\mu\nu} + h''_{\mu\nu} \Big) f_{R} + h'_{\mu\nu} f^{\;'}_{R}
    =
    0, \label{S P E 1}
\end{eqnarray}
or, we can have a more simpler form
\begin{eqnarray}
    \Box h_{\mu \nu} -\frac{f^{\,'}_{R}}{f_{R}} \partial_{y} h_{\mu \nu} = 0. \label{S P E 2}
\end{eqnarray}

By following the method given in Refs \cite{LR1999A,LR1999}, a coordinate transformation can be introduced
\begin{eqnarray}
    dz = \text{e}^{-A(y)}dy, \label{dz}
\end{eqnarray}
and the conformal flat metric can be taken:
\begin{eqnarray}
    ds^{2} = \text{e}^{2A(z)}(\eta_{\mu \nu}dx^{\mu}dx^{\nu}+dz^{2}). \label{CM}
\end{eqnarray}
By using the coordinate transformation Eq.(\ref{dz}), the perturbed Einstein equations Eq.(\ref{S P E 2}) are rewritten as:
\begin{eqnarray}
    \Big[\partial^{2}_{z} +(3\partial_{z}A +\frac{\partial_{z}f_{R}}{f_{R}})\partial_{z}
    +\hat{\Box}\Big] h_{\mu\nu} = 0.
\end{eqnarray}
By doing the KK decomposition
\begin{eqnarray}
 h_{\mu\nu}(x^{\rho},z) = \text{e}^{-\frac{3}{2}A}f_{R}^{-\frac{1}{2}}\varepsilon_{\mu\nu}(x^{\rho})\psi(z),
\end{eqnarray}
a Schr\"{o}inger-like equation for the KK modes $\psi(z)$ can be reexpressed
\begin{eqnarray}
    [-\partial^{2}_{z} + W(z)] \psi(z) = m^{2}\psi(z) \label{S Eq}
\end{eqnarray}
where the effective potential $W(z)$ is
\begin{eqnarray}
    W(z)
    &=&
    \frac{9}{4}(\partial_{z}A)^{2} + \frac{3}{2}\partial^{2}_{z}A \nonumber\\
   && + \frac{3}{2}\frac{\partial_{z}A \partial_{z}f_{R}}{f_{R}} - \frac{1}{4}\frac{(\partial_{z}f_{R})^{2}}{f^{2}_{R}}
    + \frac{\partial^{2}_{z}f_{R}}{f_{R}}. \label{Wz}
\end{eqnarray}
Above equation (\ref{Wz}) can be factorized as
\begin{eqnarray}
    QQ^{\dag}\psi(z) = m^{2}\psi(z) \label{tensor PEq}
\end{eqnarray}
with
\begin{eqnarray}
    Q
    &=&
    + \partial_{z} + \Big( \frac{3}{2}\partial_{z}A + \frac{1}{2}\frac{\partial_{z}f_{R}}{f_{R}} \Big),   \\
    Q^{\dag}
    &=& - \partial_{z} + \Big( \frac{3}{2}\partial_{z}A + \frac{1}{2}\frac{\partial_{z}f_{R}}{f_{R}} \Big),
\end{eqnarray}
which ensures that there is no gravitational mode with $m^{2}<0$ and the braneworld solutions are stable against the tensor perturbations. However, the scalar perturbations are still not clear because of the higher derivatives in the perturbation equations. It is well-known that a pure $f(R)$ gravity is conformally equivalent to a theory with a minimally coupled scalar in Einstein's gravity \cite{JS1988}. By following the method given by in Ref. \cite{YY2016}, our solutions are stable against the small linear metric perturbations, including tensor, vector, and scalar perturbations, in the Einstein frame.

For Eq.(\ref{S Eq}), the zero mode with $m = 0$ is
\begin{eqnarray}
    \psi_{(0)}(z) = N_{\text{G}}\text{e}^{\frac{3}{2}A(z)}f^{\frac{1}{2}}_{R}(z), \label{GZm}
\end{eqnarray}
where $N_{\text{G}}$ is the normalization constant, and the $\psi_{(0)}(z)$ satisfies the normalizable condition
\begin{eqnarray}
  \int^{+\infty}_{-\infty} |\psi_{(0)}(z)|^{2} dz = 1.
\end{eqnarray}

The coordinate transformation Eq.(\ref{dz}) can be rewritten as follows:
\begin{eqnarray}
 z = \int\text{e}^{-A(y)}dy = \int \cosh^{\delta}(ky)dy. \label{zy}
\end{eqnarray}
However, the above equation can not be analytically calculated for an arbitrary $\delta$, and it is also difficult to directly get the analytical expressions for the effective potential $W(z)$ and the zero mode $\psi_{(0)}(z)$ under the conformal coordinate $z$. So the solutions under the coordinate $y$ can be calculated firstly, and the numerical solutions under the conformal coordinate $z$ can also be obtained.

\begin{figure}
\begin{center}
\includegraphics[width = 0.4\textwidth]{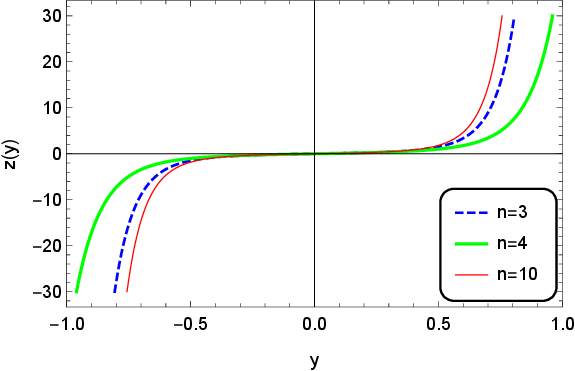}
\caption{The relation between $z$ and $y$. $n = 3$ for the dashed blue line, $n = 4$ for the green line, and $n = 10$ for the thin red line. The parameter is set to $k = 1$.}
    \label{fig_zy}
\end{center}
\end{figure}

By using the relation
\begin{eqnarray}
    \text{e}^{-A(y)}
    \xrightarrow{y \rightarrow +\infty} 2^{-\delta}\text{e}^{\,\delta ky},
\end{eqnarray}
the asymptotic behavior of the relation between $z$ and $y$ can be analyzed
\begin{eqnarray}
    z
    \xrightarrow{y \rightarrow +\infty} \frac{1}{2^{\delta}\delta k}\,\text{e}^{\,\delta ky}.
\end{eqnarray}
We can solve the asymptotic behavior of the relation between $y$ and $z$ for $z$ tends to infinity
\begin{eqnarray}
 y \xrightarrow{z \rightarrow +\infty} \frac{1}{\delta k}\ln(2^{\,\delta}\delta kz). \label{yz}
\end{eqnarray}
And the relations between $z$ and $y$ can be solved by numerical method, and shown in Fig\,\ref{fig_zy}, when $\delta$ are the determined values given by $n = 3, 4, 10$ respectively.

\begin{figure}
\begin{center}
\includegraphics[width = 0.4\textwidth]{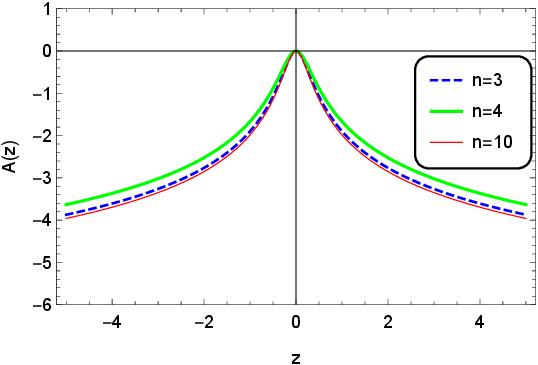}
\caption{The shapes of $A(z)$ for different solutions. $n = 3$ for the dashed blue line, $n = 4$ for the green line, and $n = 10$ for the thin red line. The parameter is set to $k = 1$.}
 \label{fig_Az}
\end{center}
\end{figure}

\begin{figure*}
\begin{center}
\subfigure[]{\label{fig_Rz1}
\includegraphics[width=0.4\textwidth]{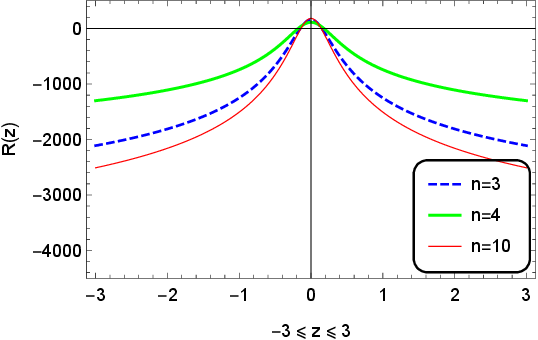}}
\subfigure[]{\label{fig_Rz2}
\includegraphics[width=0.4\textwidth]{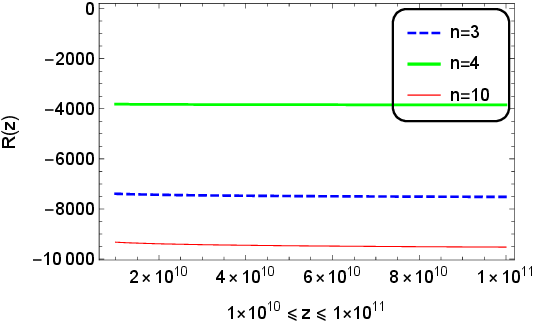}}
\end{center}\vskip -5mm
\caption{The shapes of the scalar curvature $R(z)$ for different solutions. $n = 3$ for the dashed blue line, $n = 4$ for the green line, and $n = 10$ for the thin red line. The range of the coordinate is around zero for (a), and the range of the coordinate is far away from zero for (b). The parameter is set to $k = 1$.}
    \label{fig_Rz}
\end{figure*}

Then the behavior of $A(z)$ and the scalar curvature $R(z)$ at $z = 0$ and
 $z\rightarrow \pm\infty$ can be analyzed respectively,
\begin{eqnarray}
&&
\begin{cases}
    A(z = 0) = 0 \\
    A(z \rightarrow \pm\infty) \rightarrow -\ln(\delta k|z|)
\end{cases}, \label{Az} \\
&&
\begin{cases}
    R(z = 0) = 8\delta k^{2}  \\
    R(z \rightarrow \pm\infty) \rightarrow - 20\delta^{2}k^{2} + C_{1}|z|^{-\frac{2}{\delta}}
\end{cases} \label{Rz}
\end{eqnarray}
with $C_{1} = (8\delta{k^{2}} + 20\delta^{2}k^{2}) (\delta{k})^{-\frac{2}{\delta}}$, and we can have conclusions that the warp factor $\text{e}^{2A}$ tends to $0$ and $R$ tends to a constant $-20\delta^{2}k^{2}$ for $z\rightarrow \pm\infty$.
The numerical solutions of $A(z)$ and $R(z)$ are also solved and shown in Figs.\,\ref{fig_Az} and \ref{fig_Rz}.

\begin{figure*}
\begin{center}
\subfigure[]{\label{fig Wy}
\includegraphics[width = 0.4\textwidth]{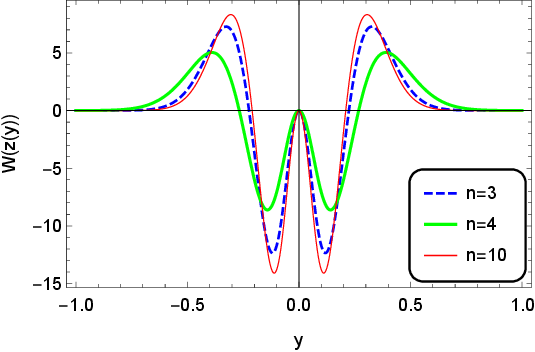}}
\subfigure[]{\label{fig Wz}
\includegraphics[width = 0.4\textwidth]{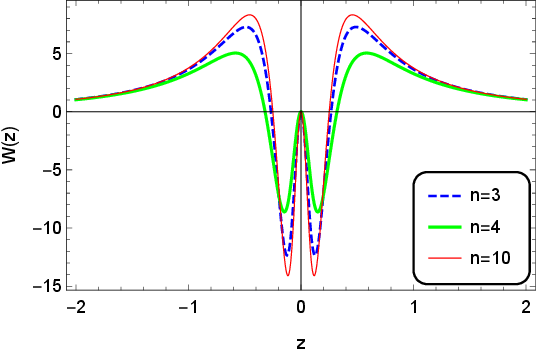}}
\end{center}\vskip -5mm
\caption{The shapes of the potential $W(z(y))$ in \ref{fig Wy} and $W(z)$ in \ref{fig Wz} for different solutions. $n = 3$ for the dashed blue line, $n = 4$ for the green line, and $n = 10$ for the thin red line. The parameter is set to $k = 1$.}
\label{fig GW}
\end{figure*}

Since there is no analytical expression for the effective potential $W(z)$ with respect to the coordinate $z$, the expression for it in the coordinate $y$ can be expressed as follows:
\begin{eqnarray}\label{Wy}
    W(z(y))
    &=&
    \frac{15}{4}\text{e}^{2A(y)}A^{'\,2}(y) +\frac{3}{2}\text{e}^{2A(y)}A''(y) \nonumber\\
    &&
    + \frac{1}{2}\text{e}^{2A(y)}\frac{4A'(y)f^{\,'}_{R} + f^{\,''}_{R}}{f_{R}} \nonumber\\
    &&
    - \frac{1}{4}\text{e}^{2A(y)}\Big(\frac{f^{\,'}_{R}}{f_{R}}\Big)^{2}.
\end{eqnarray}
 And the shapes of $W(z(y))$ are shown in Fig.\,\ref{fig Wy}. The numerically solution of $W(z)$ are shown in Fig.\,\ref{fig Wz}, which are similar to Fig.\,\ref{fig Wy}. The asymptotic behavior of these potentials tends zero at infinity. The height of the potential is related to the parameter $\delta$ decided by $n$. When $n = 10$ and $\delta = \frac{114}{5}$, the height of the potential is the tallest of all, and when $n = 4$ and $\delta = 14$, it is the lowest.

\begin{figure}
\begin{center}
\includegraphics[width = 0.4\textwidth]{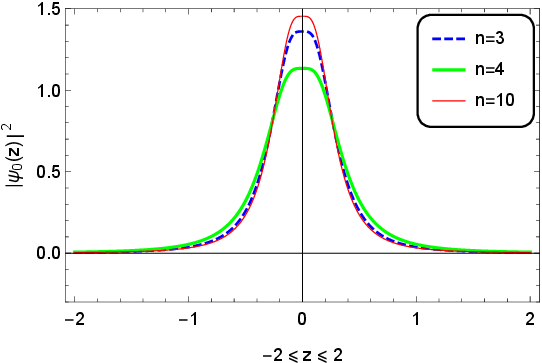}
\caption{The shapes of the zero mode for gravity $|\psi_{(0)}(z)|^{2}$ for different solutions. $n = 3$ for the dashed blue line, $n = 4$ for the green line, and $n = 10$ for the thin red line. The parameter is set to $k = 1$.}
    \label{fig_zero mode}
\end{center}
\end{figure}

This potential supports a normalizable zero mode, and the normalization constant $N_{\text{G}}$ satisfies the following condition:
\begin{eqnarray}
    1
    &=&
    \int^{+\infty}_{-\infty} |\psi_{(0)}(z)|^{2} dz
    =
    N_{\text{G}}^{2} \int^{+\infty}_{-\infty} \text{e}^{3A(z)}f_{R}(z)dz \nonumber \\
    &=&
    \frac{N^{2}_{\text{G}}}{k} \int^{+\infty}_{-\infty} \text{e}^{2A(\omega)}f_{R}(\omega)d\omega,
\end{eqnarray}
where $\omega = ky$ and $dz = \text{e}^{-A}\frac{d\omega}{k}$. For our solutions, the above integration can be done analytically. For $n = 3$, the integration gives
\begin{eqnarray}
    - \frac{N_{\text{G}}^{2}}{k} \frac{3723736645632}{245251477631875k^{2}} \Lambda = 1,
\end{eqnarray}
and $N_{\text{G}}\approx 8.115 k^{\frac{3}{2}}/\sqrt{-\Lambda}$. For $n = 4$ and $n = 10$, similarly, we can get $N_{\text{G}}\approx 6.181 k^{\frac{3}{2}}/\sqrt{-\Lambda}$, and $N_{G}\approx 8.965 k^{\frac{3}{2}}/\sqrt{-\Lambda}$ respectively.
Thus, the gravitational zero mode is normalizable and can be localized on the brane, which results in the familiar Newton's law on the brane. The shape of zero mode $|\psi_{(0)}(z)|^{2}$ is shown in Fig.\,\ref{fig_zero mode}. There is a platform for the gravitational zero mode around zero, so the zero mode is localized near zero.

\begin{figure*}
\begin{center}
\subfigure[]{\label{fig z2Wz1}
\includegraphics[width = 0.4\textwidth]{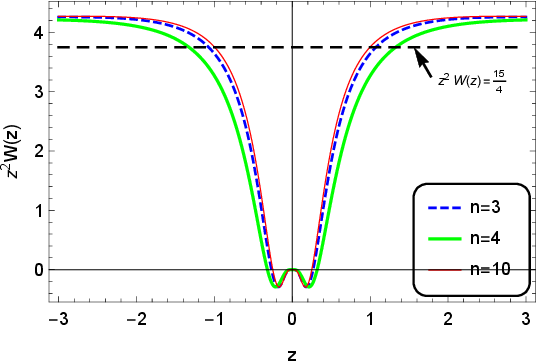}}
\subfigure[]{\label{fig z2Wz2}
\includegraphics[width = 0.4\textwidth]{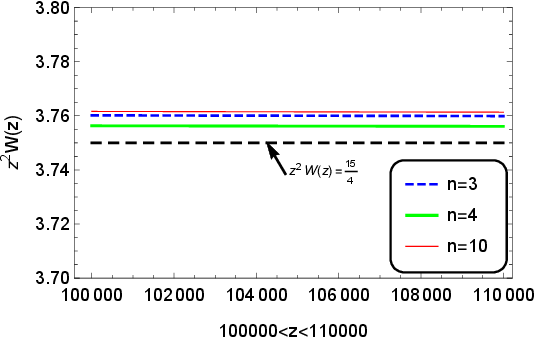}}
\end{center}\vskip -5mm
\caption{The shapes of $z^{2}W(z)$ for different solutions. $n = 3$ for the dashed blue line, $n = 4$ for the green line, and $n = 10$ for the thin red line. The parameter is set to $k = 1$.}
    \label{fig z2wz}
\end{figure*}

Starting from $m^{2}>0$, the continuum of the massive KK modes might lead to a correction to the Newtonian potential on the brane. As have been addressed in ref.\cite{CJTY2000}, if the potential
$\lim_{z\rightarrow\infty} W(z)=\frac{\beta(\beta+1)}{z^{2}}$, the massive modes will contribute a correction $\Delta U\propto 1/r^{2 \beta}$ to the Newton's law at large distance (see also \cite{DAL2009}). For our solutions, the asymptotic behavior of $z^{2}W(z)$ with $z\rightarrow\infty$ can be analytically obtained. Instead of writing down the explicit expressions, it is shown
in Fig.\,\ref{fig z2wz} that for our solutions $\lim_{z\rightarrow\infty} z^{2}W(z)=\frac{15)}{4}$, namely $\beta=3/2$. Thus, the corrections to the Newtonian potential $\Delta U\propto 1/r^{3}$ are suppressed at large $r$ for the braneworld.

Note that for KK modes with $0<m^{2}<W_{\text{max}}(z)$, there might exist some resonant KK modes which tend to plane waves when $z\rightarrow \infty$ and cannot be normallized for this type potential $W(z)$ \cite{HYZF2012}. Here $W_{\text{max}}(z)$ is the maximum of the potential. However, according to the numerical calculation, there exists no resonant state.

\section{\label{sec:scalar} Localization of Spin$-0$ scalar fields on the thick $f(R)$ branes}

In this section, we will investigate the localization of scalar field on the thick $f(R)$ branes. We start with the action of a 5D massless scalar field:
\begin{eqnarray}
    S_{0} = \int d^{5}x \sqrt{-g} \Big[-\frac{1}{2}g^{MN}\partial_{M}\Phi\partial_{N}\Phi \Big], \label{scalar action}
\end{eqnarray}
and the equation of motion from the above action(\ref{scalar action}) is read as:
\begin{eqnarray}
    \frac{1}{\sqrt{-g}}\partial_{M}(\sqrt{-g}g^{MN}\partial_{N}\Phi) = 0. \label{SM}
\end{eqnarray}
Using the conformal metric(\ref{CM}), and introducing the KK decomposition
\begin{eqnarray}
  \Phi(x^{\mu},z) = \sum_{\theta}\phi_{\theta}(x^{\mu})\chi_{\theta}(z)\text{e}^{-\frac{3}{2}A},
\end{eqnarray}
we can get the 4D Klein-Gordon equation
\begin{eqnarray}
    \frac{1}{\sqrt{-\hat{g}}}\partial_{\mu} (\sqrt{-\hat{g}}\eta^{\mu \nu}\partial_{\nu})\phi_{\theta}
    =
    m^{2}_{\theta}\phi_{\theta},
\end{eqnarray}
where $\hat{g} = \det(\eta_{\mu\nu})$ is the determinant of the 4D effective metric,
and the Schr\"{o}dinger-like equation of the scalar KK modes respect to the extra dimensional coordinate
\begin{eqnarray}
    [-\partial^{2}_{z}+V_{0}(z)]\chi_{\theta}(z) = m^{2}_{\theta}\chi_{\theta}(z), \label{scalar SEq}
\end{eqnarray}
where $m_{\theta}$ is the mass of the $\theta$-th KK mode of the scalar field, and the potential is given by:
\begin{eqnarray}
    V_{0}(z) = \frac{3}{2}\partial_{z}^{2}A +\frac{9}{4}(\partial_{z} A)^{2}. \label{SV0}
\end{eqnarray}
The scalar KK modes should satisfy the following orthogonal normalization conditions:
\begin{eqnarray}
    \int^{\infty}_{-\infty}dz\chi_{\theta}(z)\chi_{\sigma}(z) = \delta_{\theta\sigma},
\end{eqnarray}
and the action (\ref{scalar action}) of the 5D free massless scalar field turns to the 4D effective action of a massless $(m_{0} = 0)$ and a series of massive $(m_{\theta} > 0)$ scalar fields:
\begin{eqnarray}
    S_{0}
    =
    \sum_{\theta}\int d^{4}x\sqrt{-g} [- \frac{1}{2}g^{\mu\nu}\partial_{\mu}\phi_{\theta}\partial_{\nu}\phi_{\theta}
    - \frac{1}{2}m^{2}_{\theta}\phi^{2}_{\theta} ]
\end{eqnarray}
By setting $m = 0$, the scalar zero mode can be solved from Eq.(\ref{scalar SEq})
\begin{eqnarray}
    \chi_{0}(z) = N_{\text{S}}\text{e}^{\frac{3}{2}A(z)}, \label{Szm}
\end{eqnarray}
where $N_{\text{S}}$ is the normalization constant given by
\begin{eqnarray}
    1
    &=&
    \int^{+\infty}_{-\infty} |\chi_{0}(z)|^{2} dz
    =
    N_{\text{S}}^{2} \int^{+\infty}_{-\infty} \text{e}^{3A(z)}dz \nonumber \\
    &=&
    \frac{N^{2}_{\text{S}}}{k}\int^{+\infty}_{-\infty} \text{e}^{2A(\omega)}(\omega)d\omega.
\end{eqnarray}
When $n = 3$, $n = 4$, and $n = 10$, $N_{\text{S}}\approx 1.583 \sqrt{k}$, $N_{\text{S}}\approx 1.446 \sqrt{k}$, and $N_{\text{S}}\approx 1.637 \sqrt{k}$ respectively. The zero mode of scalar can be localized on the brane.

\begin{figure}
\begin{center}
\includegraphics[width = 0.35\textwidth]{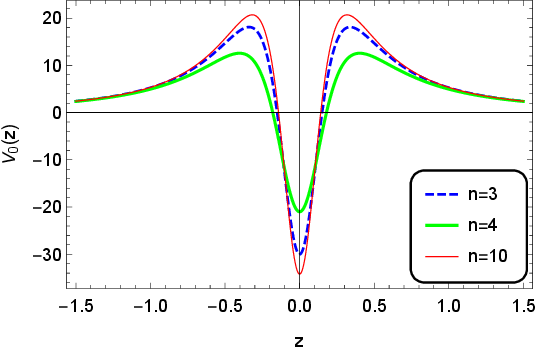}
\caption{The shapes of the effective potential $V_{0}(z)$ for different solutions. $n = 3$ for the dashed blue line, $n = 4$ for the green line, and $n = 10$ for the thin red line. The parameter is set to $k = 1$.}
    \label{fig_SVz}
\end{center}
\end{figure}

\begin{figure}
\begin{center}
\includegraphics[width = 0.35\textwidth]{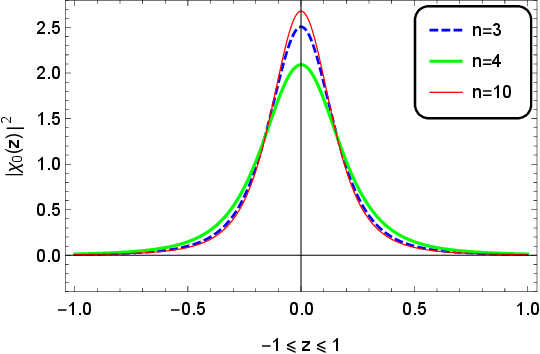}
\caption{The shapes of scalar field zero mode $|\chi_{0}(z)|^{2}$ for different solutions.
$n = 3$ for the dashed blue line, $n = 4$ for the green line, and $n = 10$ for the thin red line. The parameter is set to $k = 1$.}
    \label{fig_SZm}
\end{center}
\end{figure}

Considering Eq.(\ref{Az}), the asymptotic behavior of $V_{0}(z)$ (\ref{SV0}) and $\chi_{0}(z)$ (\ref{Szm}) can be analyzed as follows:
\begin{eqnarray}
  \begin{cases}
     V_{0}(z = 0) = -\frac{3}{2}\delta k^{2}  \\
     V_{0}(z\rightarrow\pm\infty) \rightarrow \frac{15}{4}|z|^{-2}\rightarrow 0 ,
\end{cases}
\end{eqnarray}
\begin{eqnarray}
\begin{cases}
  \chi_{0}(z = 0) = N_{\text{S}}  \\
  \chi_{0}(z \rightarrow \pm\infty) \rightarrow
  N_{\text{S}}(\delta k)^{-\frac{3}{2}} |z|^{-\frac{3}{2}}.
\end{cases} \label{SC1 behavior}
\end{eqnarray}
The numerical results of $V_{0}(z)$ and $|\chi_{0}(z)|^{2}$ are also shown in Fig.\,\ref{fig_SVz} and Fig.\,\ref{fig_SZm} respectively. We can have a conclusion that the zero mode of the scalar is localized at $z = 0$.

\section{\label{sec:vector} Localization of Spin$-1$ vector fields on the thick $f(R)$ branes}

In this section, we will investigate the localization of vector fields on the thick $f(R)$ brane. We start with a 5D gauge invariant action for a vector field coupled to the scalar curvature \cite{ZQ2018}
\begin{eqnarray}
    S_{1} = - \frac{1}{4}\int d^{5}x \sqrt{-g}g_{\text{V}}(R)g^{MN}g^{RS}F_{MR}F_{NS}, \label{vector action}
\end{eqnarray}
where $F_{MN} = \partial_{M}A_{N}-\partial_{N}A_{M}$ is the field strength tensor, and $g_{\text{V}}(R)$ is the function of scalar curvature $R$. Setting $g_{\text{V}}(R) = 1$, the action of the 5D free vector field can be recovered.
The equation of motion is read as follows:
\begin{eqnarray}
    \frac{1}{\sqrt{-g}}\partial_{M}(\sqrt{-g}g_{\text{V}}(R)g^{MN}g^{RS}F_{NS}) = 0. \label{vector motion}
\end{eqnarray}
Using the conformal metric (\ref{CM}), introducing the following general KK decomposition:
\begin{eqnarray}
    A_{\mu}(x^{\lambda},z)
    =
    \sum_{\theta}a^{(\theta)}_{\mu}(x^{\lambda}) \rho_{\theta}(z) \text{e}^{-\frac{A}{2}} (g_{\text{V}}(R))^{-\frac{1}{2}},
\end{eqnarray}
and choosing the gauge freedom $A_{5} = 0$, the vector KK modes $\rho_{\theta}(z)$ should satisfy the following Schr\"{o}dinger-like equation:
\begin{eqnarray}
    [-\partial^{2}_{z} + V_{1}(z)] \rho_{\theta}(z) = m^{2}_{\theta}\rho_{\theta}(z),
    \label{V S E}
\end{eqnarray}
where the effective potential $V_{1}(z)$ is
\begin{eqnarray}
    V_{1}(z)
    &&=
    \frac{1}{2}\partial^{2}_{z}A+\frac{1}{4}(\partial_{z}A)^{2}
      \nonumber \\
    &&~
    + \frac{\partial_{z}A\partial_{z}g_{\text{V}}(R)}{2g_{\text{V}}(R)}
    + \frac{\partial^{2}_{z}g_{\text{V}}(R)}{2g_{\text{V}}(R)}
    - \frac{(\partial_{z}g_{\text{V}}(R))^{2}}{4(g_{\text{V}}(R))^{2}}. \label{V1z}
\end{eqnarray}

The full 5D action (\ref{vector action}) can be reduced to the 4D effective action for a massless and series of massive vectors
\begin{eqnarray}
    S_{1}
    &=&
    \sum_{\theta}\int d^{4}x\sqrt{-\hat{g}}(-\frac{1}{4}\eta^{\mu \alpha} \eta^{\nu \beta}f^{(\theta)}_{\mu \nu}f^{(\theta)}_{\alpha \beta} \nonumber \\
    &&~
    - \frac{1}{2}m^{2}_{\theta}\eta^{\mu \nu}a^{(\theta)}_{\mu}a^{(\theta)}_{\nu}),
\end{eqnarray}
when integrated over the extra dimension, with the requirement that Eq. (\ref{V S E}) is satisfied and the following orthonormalization conditions are obeyed:
\begin{eqnarray}
    \int^{\infty}_{-\infty}\rho_{\theta}(z)\rho_{\sigma}(z)dz = \delta_{\theta\sigma}.
\end{eqnarray}

By setting $m = 0$, the solution of zero mode of the KK modes can be calculated:
\begin{eqnarray}
    \rho_{0}(z) = N_{\text{V}}\text{e}^{\frac{1}{2}A(z)}(g_{\text{V}}(R(z)))^{\frac{1}{2}}, \label{Vzmwct}
\end{eqnarray}
where $N_{\text{V}}$ is the normalization constant given by
\begin{eqnarray}
    1
    &=&
    \int^{+\infty}_{-\infty} |\rho_{0}(z)|^{2} dz
    =
    N_{\text{V}}^{2} \int^{+\infty}_{-\infty} \text{e}^{A(z)}g_{\text{V}}(R(z))dz \nonumber \\
    &=&
    \frac{N^{2}_{\text{V}}}{k} \int^{+\infty}_{-\infty}g_{\text{V}}(R(\omega))d\omega. \label{vector LC}
\end{eqnarray}

Next we will investigate the relation between the coupling with the gravity and the localization mechanism of the KK modes for the vector field.

\subsection{\label{sec:Vg1} Without the coupling: $g_{\text{V}}(R)=1$}

\begin{figure}
\begin{center}
\includegraphics[width = 0.35\textwidth]{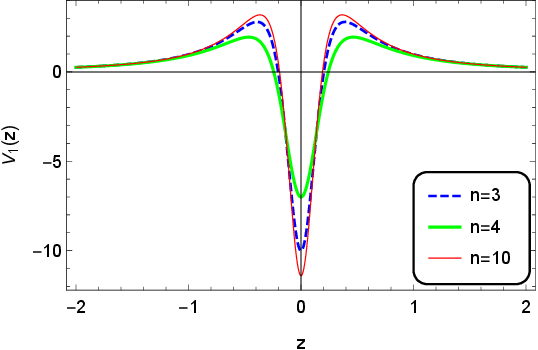}
\caption{The shapes of vector field effective potential $V_{1}(z)$ without coupling term for different solutions.
$n = 3$ for the dashed blue line, $n = 4$ for the green line, and $n = 10$ for the thin red line. The parameter is set to $k = 1$.}
    \label{fig_VV1}
\end{center}
\end{figure}

Firstly, we do not introduce the coupling between the vector field and the background spacetime, i.e.,
$g_{\text{V}}(R) = 1$. From Eq.(\ref{V1z}), the potential of the vector can be reduced to
\begin{eqnarray}
    V_{1}(z) &=& \frac{1}{2}A'' + \frac{1}{4}A'^{2}, \label{VVzq0}
\end{eqnarray}
and the asymptotic behavior of $V_{1}(z)$ at $z = 0$ and $z\rightarrow\infty$ can be analyzed as follows
\begin{eqnarray}
\begin{cases}
    V_{1}(z = 0) = \frac{1}{2} \delta k^{2} \\
    V_{1}(z\rightarrow\pm\infty) \rightarrow \frac{3}{4}z^{-2}\rightarrow 0.
\end{cases}
\end{eqnarray}
The shapes of $V_{1}(z)$ for the thick brane solutions are shown in Fig.\ref{fig_VV1} by numerical method.

The zero mode (\ref{Vzmwct}) can also be reduced to
\begin{eqnarray}
    \rho_{0}(z) = N_{\text{V}}\text{e}^{\frac{1}{2}A}, \label{Vzm}
\end{eqnarray}
and the asymptotic behavior of $\rho_{0}(z)$ has be calculated
\begin{eqnarray}
\begin{cases}
     \rho_{0}(z = 0) = N_{\text{V}}   \\
     \rho_{0}(z\rightarrow \pm\infty) \rightarrow
     \frac{1}{\sqrt{\delta k}} z^{-\frac{1}{2}}
     \rightarrow 0.
\end{cases} \label{rhoz}
\end{eqnarray}
From Eqs. (\ref{vector LC}) and (\ref{rhoz}), it is clear that the zero mode does not satisfy the normalization constant, i.e.,
\begin{eqnarray}
 \int^{\infty}_{1}|\rho_{0}(z)|^{2}dz
 \approx \int^{\infty}_{1}\frac{1}{z}dz
 \rightarrow \infty. \label{VLC}
\end{eqnarray}
Therefore the zero mode $\rho_{0}(z)$ Eq.(\ref{Vzm}) can not be localized on the brane without the coupling between the vector and the gravity. The situation is the same as the RS model.

\subsection{\label{sec:VgvR} With the coupling: $g_{\text{V}}(R) = (1+\frac{R}{20\delta^{2}k^{2}})^{q}$}

In order to localization of the zero mode of the vector field, the coupling between the vector and the gravity must be introduced, and the following simple form is choose in this paper:
\begin{eqnarray}
    g_{\text{V}}(R) = (1 + \frac{R}{20\delta^{2}k^{2}})^{q}, \label{gvR}
\end{eqnarray}
where $q$ is coupling coefficient.

\begin{figure}
\begin{center}
\includegraphics[width = 0.35\textwidth]{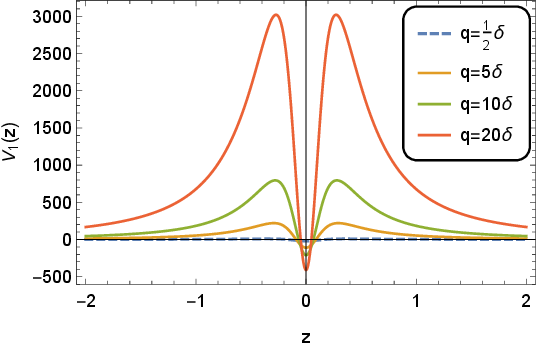}
\caption{The shapes of the effective potential for the vector fields $V_{1}(z)$ with coupling term for different values of the coupling coefficient $q$. The parameters are set to $n=3$ and $k = 1$.}
    \label{fig_VVz3q}
\end{center}
\end{figure}

\begin{figure}
\begin{center}
\includegraphics[width = 0.35\textwidth]{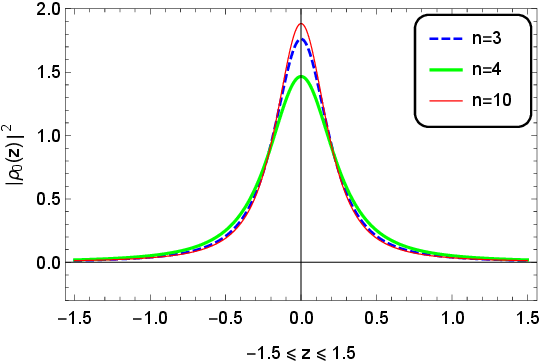}
\caption{The shapes of vector field zero modes $|\rho_{0}(z)|^{2}$ with the coupling coefficient
$q = \frac{\delta}{2}$ for different solutions.
$n = 3$ for the dashed blue line, $n = 4$ for the green line, and $n = 10$ for the thin red line. The parameter is set to $k = 1$.}
     \label{fig_Vzmq2F1}
\end{center}
\end{figure}

By using Eq.(\ref{V1z}), the effective potential can be given by
\begin{eqnarray}
 V_{1}(z)&=&\frac{A''}{2}+\frac{A'^{2}}{4}
          +\frac{q(A'R'+R'')}{40k^{2}\delta^{2}+2R} \nonumber\\
        &&~~  +\frac{q(q-2)R'^{2}}{4(20k^{2}\delta^{2}+R)^{2}},
\end{eqnarray}
and the asymptotic behaviors of $V_{1}(z)$ at zero and infinity are
\begin{eqnarray}
\begin{cases}
    V_{1}(z = 0) = -\frac{1}{2} (\delta+2q) k^{2} \\
    V_{1}(z \rightarrow \infty) \rightarrow (\frac{3}{4}+\frac{2q}{\delta}+\frac{q^{2}}{\delta^{2}})z^{-2}.
\end{cases}
\end{eqnarray}
The zero mass mode can also be given by
\begin{eqnarray}
 \rho_{0}=N_{\text{V}}\text{e}^{\frac{1}{2}A(z)}(1+\frac{R}{20k^{2}\delta^{2}})^{\frac{q}{2}},
\end{eqnarray}
and the asymptotic behaviors of the zero mode $\rho_{0}(z)$ at $z=0$ and $z\rightarrow\infty$ are
\begin{eqnarray}
\begin{cases}
    \rho_{0}(z = 0) = N_{\text{V}} \big(\frac{5\delta+2}{5\delta}\big)^{\frac{q}{2}} \\
    \rho_{0}(z \rightarrow \infty) \rightarrow N_{\text{V}} \big(\frac{5\delta+2}{5\delta}\big)^{\frac{q}{2}} (\delta{k})^{-\frac{\delta+2q}{2\delta}} z^{-\frac{\delta+2q}{2\delta}}.
\end{cases}
\end{eqnarray}
For the purpose of trapping the zero mode of the vector field on the brane, the normalization condition (\ref{vector LC}) should be satisfied, i.e.,
\begin{eqnarray}
 \int^{\infty}_{1} \rho_{0}^{2}(z\rightarrow +\infty) dz
   \approx \int^{\infty}_{1}z^{-1-2\frac{q}{\delta}} dz <\infty,
\end{eqnarray}
which require that the coupling coefficient must be positive $q > 0$. In this paper, we will investigate the effects of the coupling coefficient $q$ on the localization of the vector KK modes, so here set $q = \frac{1}{2}\delta, 5\delta, 10\delta, 20\delta$. The numerical solutions of the potential $V_{1}(z)$ for the $n=3$ thick brane are shown Fig.\,\ref{fig_VVz3q}. It is clear that the potentials have a negative well around $z=0$ and have two symmetrical barriers at both sides of the origin of the extra dimension, which increased with the coupling coefficient $q$. For the case of $n = 4$ and $n = 10$, the potentials are similar to $n = 3$. When $q = \frac{\delta}{2}$, the numerical results of the vector zero mode for $n=3$, $n=4$, and $n=10$ have been also shown in Fig.\,\ref{fig_Vzmq2F1}. Generally, this type potential implies that there may exist resonant states, which tend to plane waves when $z\rightarrow\infty$ and cannot be normalized. Following the method presented in Refs. \cite{YJZCY2009,YCLY2009,LYY2008,QJL2013}, the relative probability function of a resonance on the brane is defined as follows:
\begin{eqnarray}
    P_{\text{V}}(m^{2}) = \frac{\int^{z_{\text{b}}}_{-z_{\text{b}}}|\rho(z)|^{2}dz} {\int^{z_{\text{max}}}_{-z_{\text{max}}}|\rho(z)|^{2}dz}, \label{PV}
\end{eqnarray}
where $2 z_{\text{b}}$ is approximately the width of the thick brane, and $z_{\text{max}} = 10z_{\text{b}}$. It is clear that the KK modes are approximately taken as plane waves and the corresponding probability $P_{\text{V}}(m^{2})$ tend to $1/10$, when $m^{2}\gg V_{1}^{max}$ ($V_{1}^{max}$ is the maximum value of the corresponding potential $V_{1}$). The lifetime $\tau$ of a resonant state is $\tau \sim \Gamma^{-1}$ with $\Gamma = \delta m$ being the full width at half maximum of the peak. Equation (\ref{V S E}) can be solved by the numerical method, and we will set the coupling coefficient $q$ as different values
 $q = \frac{1}{2}\delta, 5\delta, 10\delta, 20\delta$, for each thick brane solution ($n=3$, $n=4$, and $n=10$) respectively.

 When $q=\frac{1}{2}\delta$, there is no resonant state vector KK mode for each thick brane solution. However, when $q=5\delta$, there is one resonant KK mode for each thick brane solution, and the total number of resonant KK modes increases with the coupling coefficient $q$. The mass, width ,and lifetime of the vector resonant KK modes with different values of $q$ for each thick brane solution are listed in Tables \ref{table VP3_4_10}.

\begin{figure}
\begin{center}
\subfigure[]{\label{fig_VVz3q5}
\includegraphics[width = 0.35\textwidth]{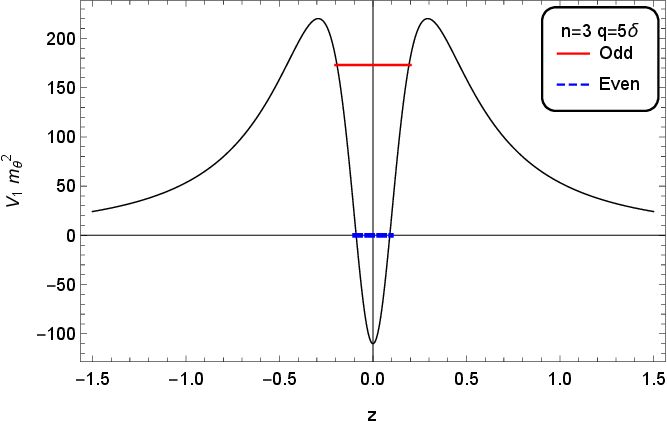}}
\subfigure[]{\label{fig_VP3q5}
\includegraphics[width = 0.35\textwidth]{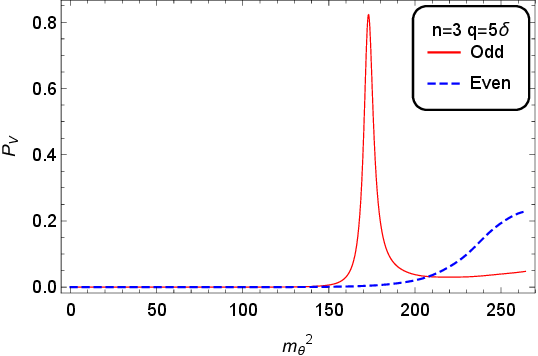}}
\subfigure[]{\label{fig_VVz3q10}
\includegraphics[width = 0.35\textwidth]{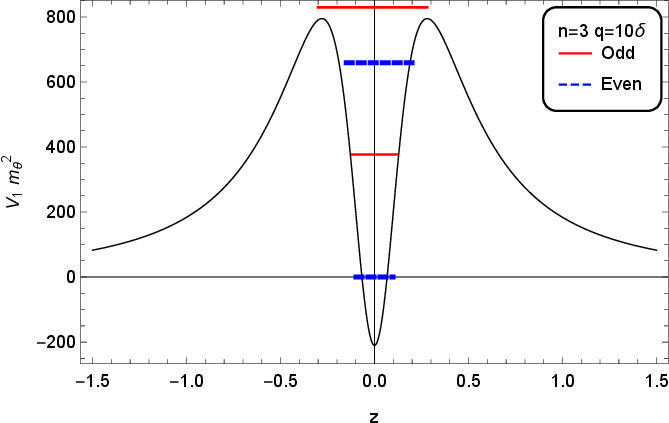}}
\subfigure[]{\label{fig_VP3q10}
\includegraphics[width = 0.35\textwidth]{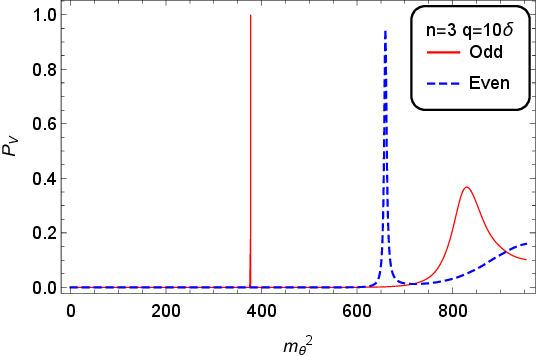}}
\subfigure[]{\label{fig_VVz3q20}
\includegraphics[width = 0.35\textwidth]{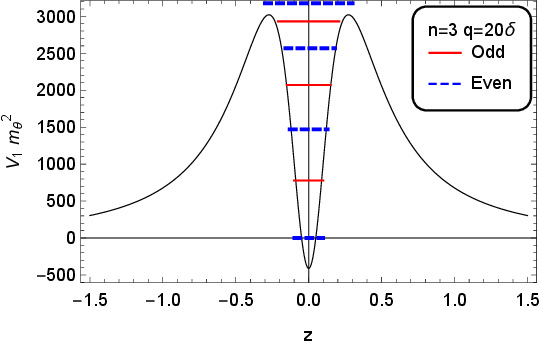}}
\subfigure[]{\label{fig_VP3q20}
\includegraphics[width = 0.35\textwidth]{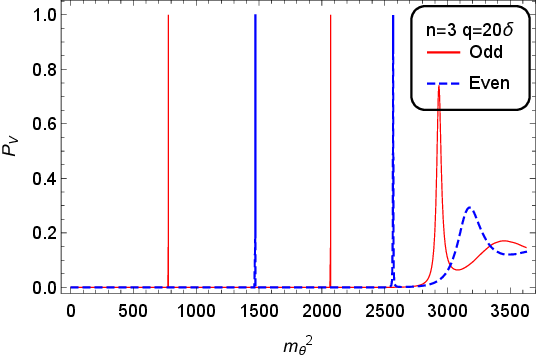}}
\end{center}\vskip -5mm
\caption{The mass spectra, the effective vector potential $V_{1}$, and corresponding relative probability $P_{\text{V}}$ with different coupling coefficient $q = 5\delta, 10\delta, 20\delta$. $V_{1}$ for the black line, the even parity KK mode for the blue line, and the odd parity KK mode for the red line. The parameters are set to $n=3$ and $k = 1$.}
    \label{fig_VP3}
\end{figure}

\begin{figure}
\begin{center}
\subfigure[]{\label{fig_V3q20wf1}
\includegraphics[width = 0.30\textwidth]{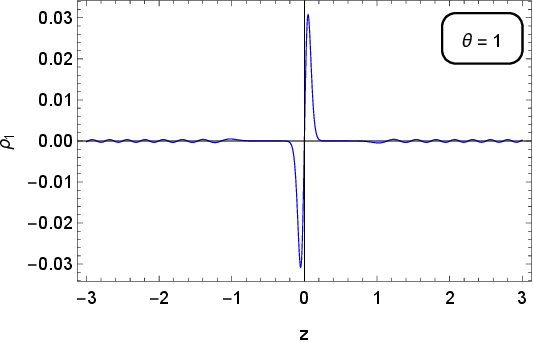}}
\subfigure[]{\label{fig_V3q20wf2}
\includegraphics[width = 0.30\textwidth]{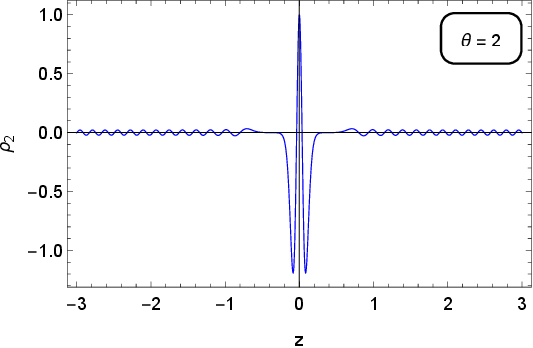}}
\subfigure[]{\label{fig_V3q20wf3}
\includegraphics[width = 0.30\textwidth]{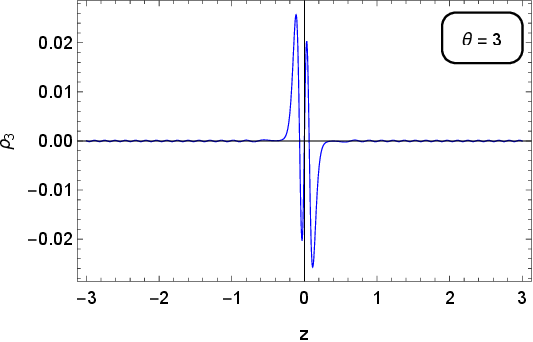}}
\subfigure[]{\label{fig_V3q20wf4}
\includegraphics[width = 0.30\textwidth]{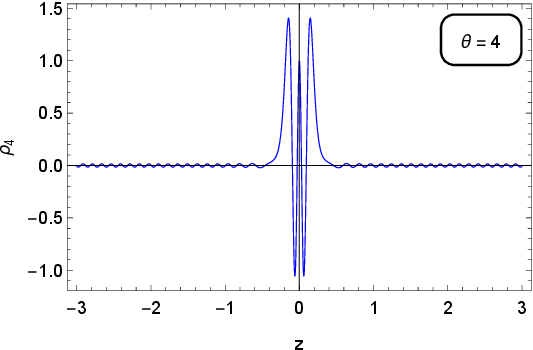}}
\subfigure[]{\label{fig_V3q20wf5}
\includegraphics[width = 0.30\textwidth]{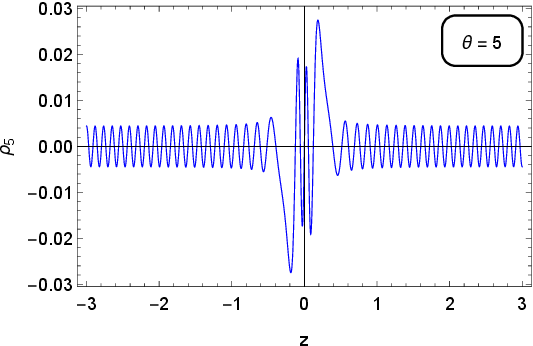}}
\subfigure[]{\label{fig_V3q20wf6}
\includegraphics[width = 0.30\textwidth]{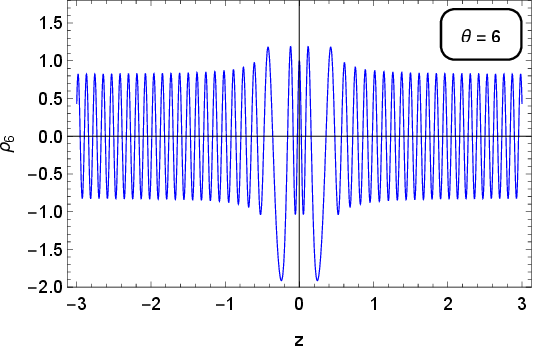}}
\end{center}\vskip -5mm
\caption{The shapes of vector resonance KK modes with the coupling coefficient $q = 20\delta$. The parameters are set to $n = 3$ and $k=1$.}
\label{fig_Vwf}
\end{figure}

 For the case of $n=3$ thick brane solution, the profiles of the relative probability $P_{\text{V}}$ corresponding to different coupling coefficient $q$ are shown in Fig.\,\ref{fig_VP3}. In these figures, each peak corresponds to a resonant state, and the corresponding mass spectra with the effective potentials are also shown in Fig.\,\ref{fig_VP3}.
 For the mass spectra of the vector KK modes, it can be seen that the ground state is zero mode (bound state), and all the massive KK modes are resonant KK modes.
 From Tables \ref{table VP3_4_10} and Fig.\,\ref{fig_VP3q20}, it is clear that there are six resonant KK modes when $q=20\delta$, and all the resonant KK modes are shown in Fig.\ref{fig_Vwf}. So we can summarize that the vector KK zero mode can be localized on the pure geometric thick $f(R)$ brane, and the massive KK modes can be quasilocalized on the brane.
 For the other cases $n = 4(\delta = 14)$ and $n = 10(\delta = \frac{114}{5})$, the situation is similar to the case of $n = 3(\delta = 20)$.
 Furthermore, the coupling function $g_{\text{V}}(R)$ can also be set as other formulations and the property of the localization will be different. And the discussion in more detail has be investigated in Ref. \cite{ZQ2018}.

\begin{table}[tbp]
\centering
\begin{tabular}{|c|c|c|c|c|c|c|c|c|}
    \hline
    $n$                 & $\delta$        & $q$                   & $V^{\text{max}}_{1}$ & $\theta$ &
    $m^{2}$             & $m$             & $\Gamma$              & $\tau$ \\
    \hline
    $3$                 & $20$            & $5\delta$             & $220.1707$           & $1$ &
    $173.1216$          & $13.1575$       & $0.2574$              & $3.8840$
    \\
    \cline{3-9}
                        &                 & $10\delta$            & $794.7859$           & $1$ &
    $377.0040$          & $19.4165$       & $6.092\times10^{-4}$  & $1.641\times10^{3}$
    \\
                        &                 &                       &                      & $2$ &
    $659.3324$          & $25.6774$       & $0.1214$              & $8.2327$
    \\
                        &                 &                       &                      & $3$ &
    $829.4739$          & $28.8005$       & $1.4998$              & $0.6667$
    \\
    \cline{3-9}
                        &                 & $20\delta$            & $3.020\times10^{3}$  & $1$ &
    $778.1259$          & $27.8949$       & $1.431\times10^{-11}$ & $6.984\times10^{10}$
    \\
                        &                 &                       &                      & $2$ &
    $1.470\times10^{3}$ & $38.3434$       & $6.533\times10^{-7}$  & $1.530\times10^{6}$
    \\
                        &                 &                       &                      & $3$ &
    $2.070\times10^{3}$ & $45.5006$       & $3.880\times10^{-4}$  & $2.577\times10^{3}$
    \\
                        &                 &                       &                      & $4$ &
    $2.567\times10^{3}$ & $50.6670$       & $0.0273$              & $36.5568$
    \\
                        &                 &                       &                      & $5$ &
    $2.931\times10^{3}$ & $54.1458$       & $0.4101$              & $2.4379$
    \\
                        &                          &                       &                      & $6$ &
    $3.176\times10^{3}$ & $56.3604$                & $2.4695$              & $0.4049$
    \\
    \hline
    $4$                 & $14$                     & $5\delta$             & $152.5827$           & $1$ &
    $120.7882$          & $10.9903$                & $0.2455$              & $4.0724$
    \\
    \cline{3-9}
                        &                          & $10\delta$            & $550.7041$           & $1$ &
    $263.5687$          & $16.2348$                & $6,529\times10^{-4}$  & $1.531\times10^{3}$
    \\
                        &                          &                       &                      & $2$ &
    $459.9604$          & $21.4466$                & $0.1200$              & $8.3271$
    \\
                        &                          &                       &                      & $3$ &
    $581.1552$          & $24.1071$                & $1.5808$              & $0.6325$
    \\
    \cline{3-9}
                        &                          & $20\delta$            & $2.092\times10^{3}$  & $1$ &
    $544.3748$          & $23.3318$                & $1.988\times10^{-11}$ & $5.028\times10^{10}$
    \\
                        &                          &                       &                      & $2$ &
    $1.027\times10^{3}$ & $32.0602$                & $8.152\times10^{-7}$  & $1.226\times10^{6}$
    \\
                        &                          &                       &                      & $3$ &
    $1.446\times10^{3}$ & $38.0281$                & $4.501\times10^{-4}$  & $2.221\times10^{3}$
    \\
                        &                          &                       &                      & $4$ &
    $1.790\times10^{3}$ & $42.3200$                & $0.0295$              & $33.8816$
    \\
                        &                          &                       &                      & $5$ &
    $2.042\times10^{3}$ & $45.1897$                & $0.4127$              & $2.4229$
    \\
                        &                          &                       &                      & $6$ &
    $2.230\times10^{3}$ & $47.2321$                & $3.6694$              & $0.2725$
    \\
    \hline
    $10$                   & $\frac{114}{5}$       & $5\delta$             & $251.7237$           & $1$ &
    $197.5419$             & $14.0549$             & $0.2653$              & $3.7686$
    \\
    \cline{3-9}
                           &                       & $10\delta$            & $908.7305$           & $1$ &
    $429.9404$             & $20.7350$             & $6.001\times10^{-4}$  & $1.666\times10^{3}$
    \\
                           &                       &                       &                      & $2$ &
    $752.3325$             & $27.4286$             & $0.1222$              & $8.1773$
    \\
                           &                       &                       &                      & $3$ &
    $945.6910$             & $30.7520$             & $1.5185$              & $0.6585$
    \\
    \cline{3-9}
                           &                       & $20\delta$            & $3.453\times10^{3}$  & $1$ &
    $887.2097$             & $29.7860$             & $1.301\times10^{-11}$ & $7.686\times10^{10}$
    \\
                           &                       &                       &                      & $2$ &
    $1.676\times10^{3}$    & $40.9470$             & $6.170\times10^{-7}$  & $1.620\times10^{6}$
    \\
                           &                       &                       &                      & $3$ &
    $2.361\times10^{3}$    & $48.5960$             & $3.745\times10^{-4}$  & $2.670\times10^{3}$
    \\
                           &                       &                       &                      & $4$ &
    $2.929\times10^{3}$    & $54.1234$             & $0.0268$ & $37.2132$
    \\
                           &                       &                       &                      & $5$ &
    $3.346\times10^{3}$    & $57.8474$             & $0.4162$ & $2.4025$
    \\
                           &                       &                       &                      & $6$ &
    $3.615\times10^{3}$    & $60.1269$             & $2.5566$ & $0.3911$
    \\
    \hline
\end{tabular}
\caption{The mass, width, and lifetime of resonant KK modes of the vector with $k = 1$. }
    \label{table VP3_4_10}
\end{table}

\section{Localization of Spin$-1/2$ fermion fields on the thick $f(R)$ branes}
\label{sec:fermion}

In this section, the localization of Spin$-1/2$ fermion field on the thick $f(R)$ branes will be investigated, and the 5D Dirac action of fermion can be expressed as
\begin{eqnarray}
    S_{\frac{1}{2}}
    &=&
    \int d^{5}x\sqrt{-g} [\bar{\Psi}\Gamma^{M} (\partial_{M}+\omega_{M}) \Psi \nonumber \\
    &&~
    - \eta\bar{\Psi}\Gamma^{M}\partial_{M}g_{\text{f}}(R)\Psi], \label{fermion action}
\end{eqnarray}
where $\omega_{M}$ is the spin connection defined as
\begin{eqnarray}
\omega_{M} = \frac{1}{4}\omega^{\bar{M}\bar{N}}_{M}\Gamma_{\bar{M}}\Gamma_{\bar{N}}
\end{eqnarray}
with
\begin{eqnarray}
    \omega^{\bar{M}\bar{N}}_{M}
    &=&
    \frac{1}{2}E^{N\bar{M}} (\partial_{M}E^{\bar{N}}_{N}-\partial_{N}E^{\bar{N}}_{M}) \nonumber \\
    &&~
    - \frac{1}{2}E^{N\bar{N}}(\partial_{M}E^{\bar{M}}_{N}-\partial_{N}E^{\bar{M}}_{M}) \nonumber \\
    &&~
    -\frac{1}{2}E^{P\bar{M}}E^{Q\bar{N}}E^{\bar{R}}_{M}
    (\partial_{P}E_{Q\bar{R}}-\partial_{Q}E_{P\bar{R}}).
\end{eqnarray}
The letters with barrier $\bar{M},\bar{N}$ are the five dimensional local Lorentz indices and the vielbein $E^{M}_{\bar{M}}$ satisfies
$E^{M}_{\bar{M}}E^{N}_{\bar{N}}\eta^{\bar{M}\bar{N}} = g^{MN}$.
The relation between the gamma matrices $\Gamma^{M}$ and
$\Gamma^{\bar{M}} = (\Gamma^{\bar{\mu}},\Gamma^{\bar{5}}) = (\gamma^{\bar{\mu}},\gamma^{\bar{5}})$
is given by
$\Gamma^{M} = E^{M}_{\bar{m}}\Gamma^{\bar{M}}$.

Here the coupling between Dirac fermion and background spacetime \cite{YYWY2017}
 $- \eta\bar{\Psi}\Gamma^{M}\partial_{M}g_{\text{f}}(R)\Psi$ with $\eta$ coupling coefficient is introduced. Here we assume that the coupling coefficient $\eta$ is positive.

Considering the conformally flat metric Eq.(\ref{CM}), the component of the spin connection is given by $\omega_{\mu} = \frac{1}{2}(\partial_{z}A(z))\gamma_{\mu}\gamma_{5}$
and $\omega_{5} = 0$.
The equation of motion of the 5D Dirac fermion can be derived as
\begin{eqnarray}
    [\gamma^{\mu}\partial_{\mu} + \gamma^{5}(\partial_{z} + 2\partial_{z}A(z)) + \eta\partial_{z}g_{\text{f}}(R)]\Psi = 0. \label{fermion motion}
\end{eqnarray}

The chiral decomposition for $\Psi(x,z)$ can be introduced,
\begin{eqnarray}
    \Psi(x,z) = \text{e}^{-2A(z)} \sum_{\theta} [\psi_{L_{\theta}}(x)L_{\theta}(z) + \psi_{R_{\theta}}(x) R_{\theta}(z)], \label{Psi xz}
\end{eqnarray}
where
$\Psi_{L_{\theta}} = -\gamma^{5} \Psi_{L_{\theta}}$
and
$\Psi_{R_{\theta}} = \gamma^{5} \Psi_{R_{\theta}}$
are the left- and right-chiral components of the 4D Dirac fermion field, respectively.

By takeing the following orthonormalization conditions for the KK modes $L_{\theta}$ and $R_{\theta}$
\begin{eqnarray}
    \int^{\infty}_{-\infty} L_{\theta} L_{\sigma} dz &=& \delta_{\theta\sigma}, \\
    \int^{\infty}_{-\infty} R_{\theta} R_{\sigma} dz &=& \delta_{\theta\sigma}, \\
    \int^{\infty}_{-\infty} L_{\theta} R_{\theta} dz &=& 0,
\end{eqnarray}
one can take the effective action of the 4D massless and massive Dirac fermions from the 5D Dirac action (\ref{fermion action})
\begin{eqnarray}
    S_{\frac{1}{2}}
    =
    \sum_{\theta}\int d^{4}x \sqrt{-\hat{g}} \bar{\psi_{\theta}}
    [\gamma^{\mu} (\partial_{\mu} + \hat{\omega}_{\mu}) - m_{\theta}] \psi_{\theta}.
\end{eqnarray}

By introducing the chiral decomposition Eq.(\ref{Psi xz}), the Schr\"{o}dinger-like equations of motion for the left- and right-chiral fermion KK modes $L_{\theta}(z)$ and $R_{\theta}(z)$ can be obtained:
\begin{subequations} \label{fermion SEq}
\begin{eqnarray}
    && [-\partial^{2}_{z} + V_{\text{L}}(z)] L_{\theta}(z) = m^{2}_{\theta} L_{\theta}(z), \label{fermion SEqL}\\
    && [-\partial^{2}_{z} + V_{\text{R}}(z)] R_{\theta}(z) = m^{2}_{\theta} R_{\theta}(z), \label{fermion SEqR}
\end{eqnarray}
\end{subequations}
where the effective potentials $V_{\text{L,R}}(z)$ of the fermion KK modes are read as
\begin{subequations} \label{fermion Vz}
\begin{eqnarray}
    V_{\text{L}}(z) = (\eta \partial_{z} g_{\text{f}}(R))^{2} +\eta \partial^{2}_{z}g_{\text{f}}(R), \label{fermion VL} \\
    V_{\text{R}}(z) = (\eta \partial_{z} g_{\text{f}}(R))^{2} -\eta \partial^{2}_{z}g_{\text{f}}(R). \label{fermion VR}
\end{eqnarray}
\end{subequations}
By setting $m_{\theta}=0$ in Eq. (\ref{fermion SEq}), the solution of zero modes $L_{0}$ and $R_{0}$ can be obtained
\begin{subequations} \label{fermion ZM}
\begin{eqnarray}
    && L_{0}
    \propto \text{e}^{\int^{z}_{0} dz \eta \partial_{z}g_{\text{f}}(R)}
    = \text{e}^{\eta g_{\text{f}}(R(z))} \label{fermion ZML} \\
    &&
    R_{0}
    \propto  \text{e}^{-\int^{z}_{0} dz\eta \partial_{z}g_{\text{f}}(R)}
    = \text{e}^{-\eta g_{\text{f}}(R(z))}. \label{fermion ZMR}
\end{eqnarray}
\end{subequations}

From the above relations (\ref{fermion ZM}), it is impossible to make both massless left- and right-chiral KK modes to be localized on the brane at the same time, since when one is normalizable, the other one is not.

From Eqs. (\ref{fermion SEq}) and (\ref{fermion Vz}), it is clear that, if we do not introduce the coupling in the action (\ref{fermion action}), i.e., $\eta=0$, the effective potentials for left- and right-chiral KK modes $V_{\text{L,R}}(z)=0$ and both left- and right-chiral fermions can not be localized on the thick brane, so the coupling term must be introduced. Moreover, since $R$ is even function of $z$, $V_{\text{L,R}}(z)$ are naturally Z$_2$ even with respect to $z$.
Here we set a simple formulation:
\begin{eqnarray}\label{gfR}
    g_{\text{f}}(R) = - \frac{20\delta^{2}k^{2} + 8\delta k^{2}}{R + 20\delta^{2}k^{2}},
\end{eqnarray}

By using Eqs.(\ref{Az},\ref{Rz}) and (\ref{gfR}), the asymptotic behaviors of $V_{\text{L,R}}(z)$ at $z=0$ and $z\rightarrow\infty$ are as follows
\begin{eqnarray}
\begin{cases}
    V_{\text{L}}(z = 0) = -2\eta k^{2}, \\
    V_{\text{L}}(z \rightarrow \infty) \rightarrow \frac{2\eta}{\delta^{2}}(1-\delta)z^{-2},
\end{cases}  \\
\begin{cases}
    V_{\text{R}}(z = 0) = 2\eta k^{2}, \\
    V_{\text{R}}(z \rightarrow \infty) \rightarrow \frac{2\eta}{\delta^{2}} (\delta - 1) z^{-2}.
\end{cases}
\end{eqnarray}
Considering the coupling formulation (\ref{gfR}) and Eq.(\ref{fermion ZM}), for the positive coupling coefficient $\eta$ and positive parameter $\delta$ and $k$, only left-chiral fermion zero mode may be localized on the brane.
The asymptotic behavior of $L_{0}(z)$ is also given by
\begin{eqnarray}
    \begin{cases}
    L_{0}(z = 0) = N_{L} \text{e}^{-\eta}, \\
    L_{0}(z \rightarrow \infty)
    \rightarrow N_{L} \text{e}^{-\eta C_{L}
                    |z|^{s}},
\end{cases}
\end{eqnarray}
where $N_{L}$ is a constant, $C_{L}=(\delta k)^{\frac{\delta}{2}}>0$ and $s=\frac{\delta}{2}\geq\frac{\delta_{\text{min}}}{2}=7$ are positive constants, respectively.
It is easy to see that the normalization condition of the left-chiral fermion zero mode is satisfied, and it can be localized on the brane.

Next, the effect of the coupling between fermion and background spacetime for the property of the localization will be investigated, so the coupling coefficient $\eta$ will be taken different values: $\eta = 1, 50, 100$. This type potential implies that resonant left- and right-chiral KK modes may exist. Mimic to the case of vector, the fermion relative probabilities for finding the left- and right-chiral fermion resonant states with mass $m$ can be defined as:
\begin{eqnarray}
    P_{\text{L,R}}(m^{2}) = \frac{\int_{-z_{b}}^{z_{b}}|L,R(z)|^{2}dz}{\int_{-z_{max}}^{z_{max}}|L,R(z)|^{2}dz}.
\end{eqnarray}

\begin{figure*}
\begin{center}
\subfigure[]{\label{fig FVLz3et1}
\includegraphics[width = 0.35\textwidth]{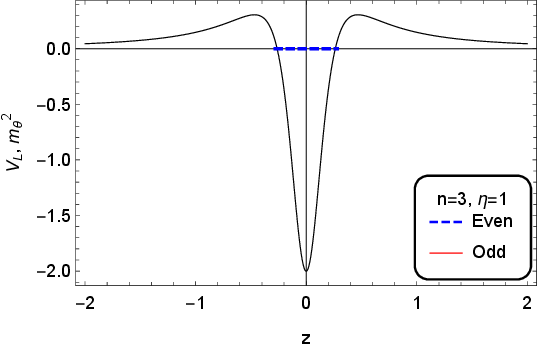}}
\subfigure[]{\label{fig FVRz3et1}
\includegraphics[width = 0.35\textwidth]{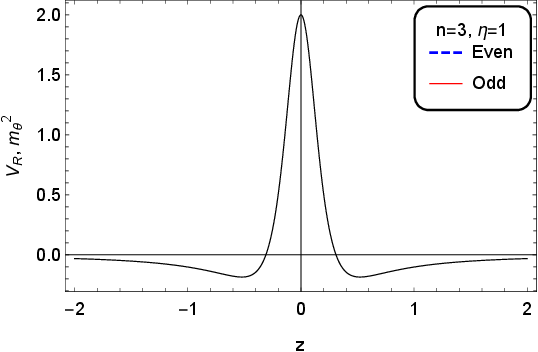}}
\subfigure[]{\label{fig FVLz3et50}
\includegraphics[width = 0.35\textwidth]{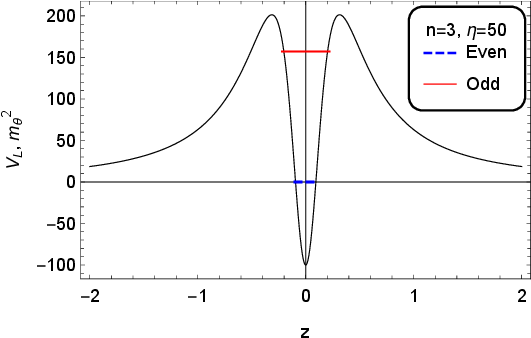}}
\subfigure[]{\label{fig FVRz3et50}
\includegraphics[width = 0.35\textwidth]{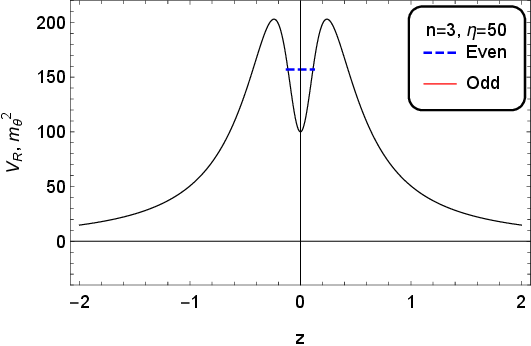}}
\subfigure[]{\label{fig FVLz3et100}
\includegraphics[width = 0.35\textwidth]{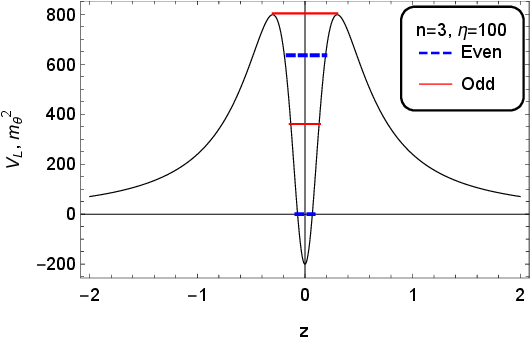}}
\subfigure[]{\label{fig FVRz3et100}
\includegraphics[width = 0.35\textwidth]{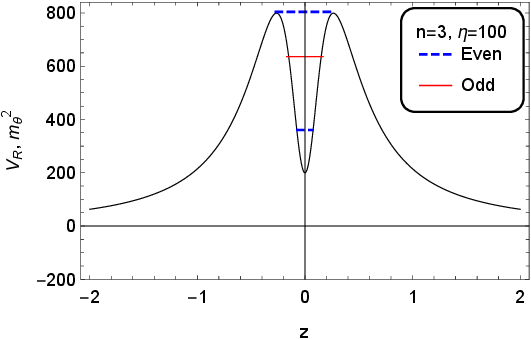}}
\end{center}\vskip -5mm
\caption{The mass spectra and the effective potentials of the left- and right-chiral fermions with different coupling coefficient $\eta=1$, $\eta=50$, and $\eta=100$. The parameters are set to $n = 3$ and $k = 1$.}
    \label{fig FVLRz3}
\end{figure*}

\begin{figure*}
\begin{center}
\subfigure[]{\label{fig FPL3et100}
\includegraphics[width = 0.35\textwidth]{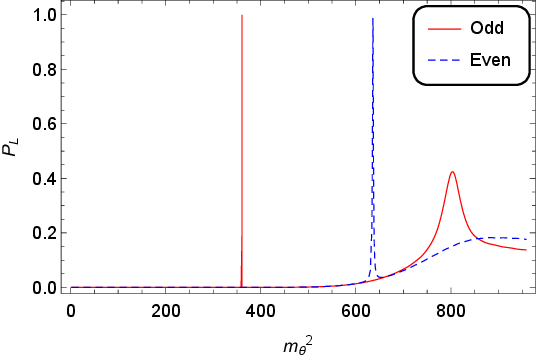}}
\subfigure[]{\label{fig FPR3et100}
\includegraphics[width = 0.35\textwidth]{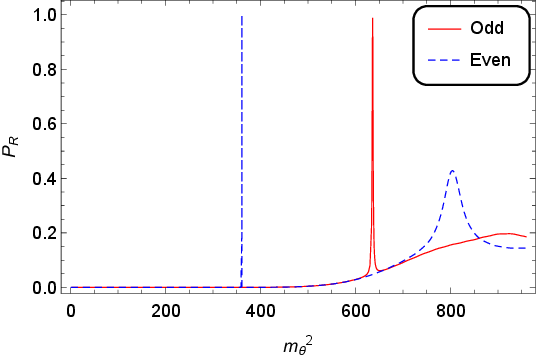}}
\end{center}\vskip -5mm
\caption{The shapes of the relative probability $P_{L}(m^{2})$ for the left-chiral fermion and $P_{R}(m^{2})$ for right-chiral fermion resonant. The parameters are set to $\eta = 100$, $n = 3$ and $k = 1$.}
    \label{fig FPLR3et100}
\end{figure*}

\begin{figure*}
\begin{center}
\subfigure[]{\label{fig FL3et100Wf0}
\includegraphics[width = 0.23\textwidth]{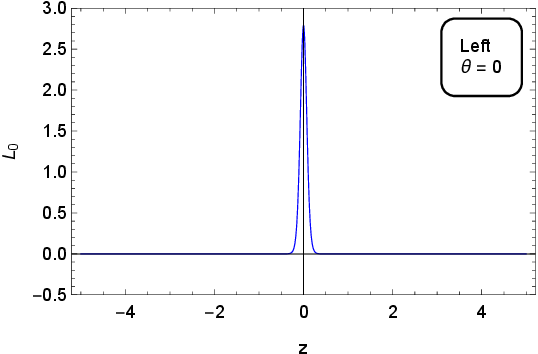}}
\subfigure[]{\label{fig FR3et100Wf1}
\includegraphics[width = 0.23\textwidth]{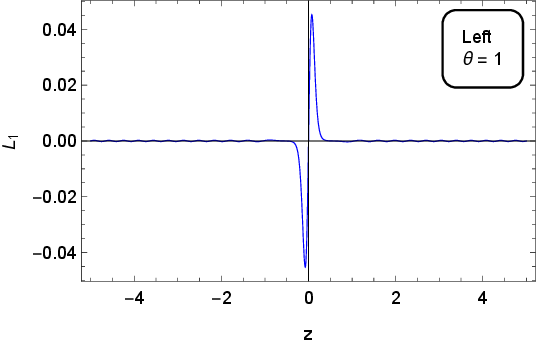}}
\subfigure[]{\label{fig FL3et100Wf2}
\includegraphics[width = 0.23\textwidth]{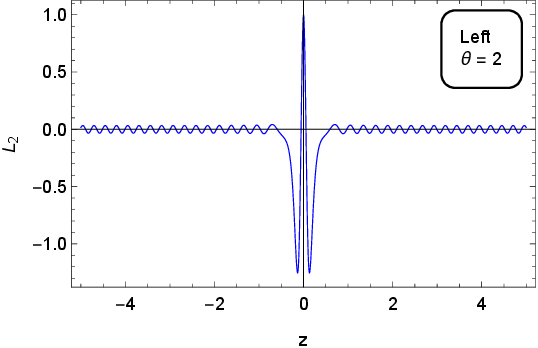}}
\subfigure[]{\label{fig FL3et100Wf3}
\includegraphics[width = 0.23\textwidth]{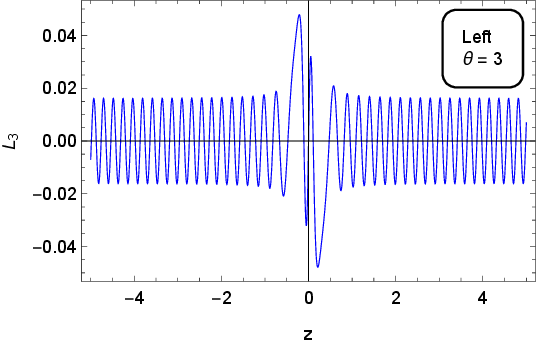}}
\end{center}\vskip -5mm
\caption{The shapes of the left-chiral fermion KK modes, which is set to $\eta = 100$, $n = 3$ and $k = 1$.}
    \label{fig FL3et100Wf}
\end{figure*}

\begin{figure*}
\begin{center}
\subfigure[]{\label{fig FR3et100Wf1}
\includegraphics[width = 0.23\textwidth]{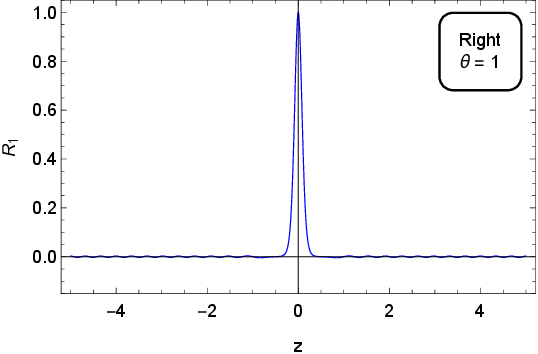}}
\subfigure[]{\label{fig FR3et100Wf2}
\includegraphics[width = 0.23\textwidth]{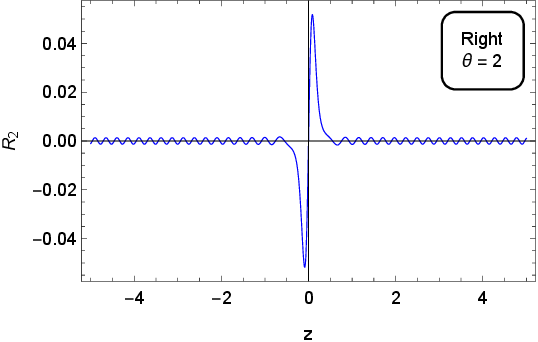}}
\subfigure[]{\label{fig FR3et100Wf3}
\includegraphics[width = 0.23\textwidth]{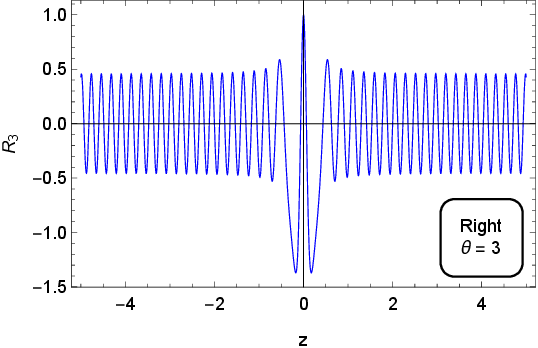}}
\end{center}\vskip -5mm
\caption{The shapes of the right-chiral fermion KK modes, which is set to $\eta = 100$, $n = 3$ and $k = 1$.}
    \label{fig FR3et100Wf}
\end{figure*}

For the case of $n=3$, when the coupling coefficient $\eta=1$, the effective potentials and the mass spectra for the left- and right-chiral fermion KK modes are shown in
Figs. \ref{fig FVLz3et1} and \ref{fig FVRz3et1}. Only the left-chiral fermion zero mode (bound state) can be localized on the brane. When the coupling coefficient $\eta=50$, the effective potentials and the mass spectra for the left- and right-chiral fermion KK modes are shown in
Figs. \ref{fig FVLz3et50} and \ref{fig FVRz3et50}. For the left-chiral fermion KK modes, there is only one bound zero mode and one massive resonant KK mode, which is an even-parity. However, for the right-chiral fermion KK mode, there is only one massive resonant KK mode, which is an odd-parity. Both the mass of the left- and right-chiral fermion KK modes are the same. In fact, these conclusions are originated from the coupled equations of the left- and right-chiral fermions. When $\eta=100$, the effective potentials and the mass spectra for the left- and right-chiral fermion KK modes are shown in Figs. \ref{fig FVLz3et100} and \ref{fig FVRz3et100}.
The profiles of the relative probability $P_{L,R}(m^{2})$ are shown in Figs.\,\ref{fig FPL3et100} and \ref{fig FPR3et100}, respectively. And the left- and right-chiral fermion KK modes are shown in Figs.\,\ref{fig FL3et100Wf} and \ref{fig FR3et100Wf}. So we can summarize that the 4D massless left-chiral fermion can be localized on the brane, and the 4D massive Dirac fermions can also be quasilocalized on the brane, which consist of the pairs of coupled left- and right-chiral KK modes with different parities. The total number of resonant KK modes increases with the coupling coefficient $\eta$.

For the cases of $n=4$ and $n=10$, the situation is similar to the case of $n=3$, and the mass, width, and lifetime of the left- and right-chiral fermion KK resonant modes are listed in Table \ref{table FP3_4_10}.

Furthermore, the coupling function $g_{\text{f}}(R)$ can also be set as other formulations and the property of the localization will be different. And the discussion in more detail has be investigated in Ref. \cite{YYWY2017}.

\begin{table}[tbp]
\begin{tabular}{|c|c|c|c|c|c|c|c|c|}
    \hline
    $n$        & $\eta$    & Chrial    & Height of $V_{L,R}$ & $\theta$ &
    $m^{2}$    & $m$       & $\Gamma$  & $\tau$
    \\
    \hline
    $3$        & $50$      & Left      & $201.3102$          & $1$ &
    $157.0089$ & $12.5303$ & $0.1732$  & $5.7749$
    \\
               &            & Right    & $202.9641$ &           $1$ &
    $156.9938$ &  $12.5297$ & $0.1801$ & $5.5527$
    \\
    \cline{2-9}
     &            $100$ &      Left &                $798.2219$ &           $1$ &
    $360.5018$ &  $18.9869$ & $9.592\times10^{-5}$ & $1.042\times10^{4}$
    \\
    &             &           &                      &                     $2$ &
    $635.9341$ &  $25.2177$ & $0.0464$ &             $21.5500$
    \\
     &             &           &                      &                     $3$ &
    $803.5692$ &  $28.3473$ & $1.2032$ &             $0.8311$
    \\
    \cline{3-9}
     &             &          Right &                $798.9972$ &           $1$ &
    $360.5020$ &  $18.9868$ & $9.555\times10^{-5}$ & $1.046\times10^{4}$
    \\
     &             &           &                      &                     $2$ &
    $635.9326$ &  $25.2177$ & $0.0470$ &             $21.2549$
    \\
     &             &           &                      &                     $3$ &
    $803.7101$ &  $28.3497$ & $1.3213$ &             $0.7568$
    \\
    \hline
    $4$ &                 $50$ &      Left &                 $296.0509$ &           $1$ &
    $172.5166$ &          $13.1345$ & $0.0057$ &             $173.6421$
    \\
     &                     &           &                      &                     $2$ &
    $281.7977$ &          $16.7868$ & $0.4054$ &             $2.4664$
    \\
    \cline{3-9}
     &                     &          Right &                $296.7473$ &           $1$ &
    $172.5166$ &          $13.1345$ & $0.0058$ &             $172.3199$
    \\
     &                     &           &                      &                     $2$ &
    $281.6335$ &          $16.7819$ & $0.4227$ &             $2.3652$
    \\
    \cline{2-9}
     &                    $100$ &     Left &                 $1.179\times10^{3}$ &  $1$ &
    $373.4193$ &          $19.3240$ & $1.844\times10^{-9}$ & $5.421\times10^{8}$
    \\
     &                     &           &                      &                     $2$ &
    $691.9976$ &          $26.3058$ & $4.681\times10^{-5}$ & $2.136\time10^{4}$
    \\
     &                     &           &                      &                     $3$ &
    $950.5699$ &          $30.8313$ & $0.0113$ &             $88.4840$
    \\
     &                     &           &                      &                     $4$ &
    $1.134\times10^{3}$ & $33.6808$ & $0.2806$ &             $3.5632$
    \\
    \cline{3-9}
     &                     &          Right &                $1.179\times10^{3}$ &  $1$ &
    $373.4196$ &          $19.3240$ & $1.868\times10^{-9}$ & $5.353\times10^{8}$
    \\
     &                     &           &                      &                     $2$ &
    $691.9974$ &          $26.3058$ & $4.647\times10^{-5}$ & $2.151\times10^{4}$
    \\
      &                     &           &                      &                     $3$ &
    $950.5701$ &          $30.8313$ & $0.0114$ &             $87.4471$
    \\
     &                     &           &                      &                     $4$ &
    $1.134\times10^{3}$ & $33.6800$ & $0.2929$ &             $3.4134$
    \\
    \hline
    $10$ &        $50$ &      Left &     $175.4931$ &           $1$ &
    $149.2316$ &  $12.2160$ & $0.3738$ & $2.6751$
    \\
     &             &          Right &    $177.7526$ &           $1$ &
    $149.0260$ &  $12.2076$ & $0.4045$ & $2.4718$
    \\
   \cline{2-9}
     &            $100$ &     Left &     $694.0240$ &           $1$ &
    $354.2681$ &  $18.8220$ & $0.0010$ & $929.1899$
    \\
     &             &           &          &                     $2$ &
    $606.2331$ &  $24.6218$ & $0.1903$ & $5.2540$
    \\
    \cline{3-9}
     &             &          Right &    $695.0366$ &           $1$ &
    $354.2683$ &  $18.8220$ & $0.0011$ & $909.9818$
    \\
     &             &           &          &                     $2$ &
    $606.3846$ &  $24.6248$ & $0.1982$ & $5.0438$
    \\
    \hline
\end{tabular}
\centering
\caption{The mass, width, and lifetime of resonant KK modes of the fermions with the parameter $k=1$.}
    \label{table FP3_4_10}
\end{table}

\section{Conclusion and discussion}
\label{sec:conclusion}

In this paper, we investigate a pure geometric thick $\mathcal{M}_{4}$ brane, embedded in a AdS$_5$ spacetime, in general $f(R)$ gravity theory. Here, the form of $f(R)$ is set as
$f(R) = \sum_{i = 1}^{n} a_{i}\;R^{\,i} +\Lambda$, and any background scalar has not been introduced. For the pure geometric thick $\mathcal{M}_{4}$ brane, the parameter $n$ must be
satisfied $n\geq 3$, and for a certainty value of $n$, a thick brane solution can be obtained. In this paper, the solutions of the $\mathcal{M}_{4}$ brane for $n=3$, $n=4$, and $n=10$, have been studied, and found that the solutions are stable against tensor perturbations.  Moreover, we investigate the localization of gravity and various bulk matter fields on the branes.

For the gravity, it is found that the gravitational zero mode can be localized near zero. All the massive modes are continuous-spectrum wave functions and can not be localized on the brane. For the scalar field, the zero mode is localized on the zero. For the vector field, if the coupling between vector and background spacetime is not introduced, the vector zero mode can not be localized on the brane, with the same case of RS brane. However, by introducing the coupling, the zero mode can be localized on the brane. For a large coupling coefficient, there exist vector resonant states, and the number of the resonances increases with the coupling coefficient. For the spin-$1/2$ fermion field, in order to localize the fermion zero mode, the coupling between the fermion and the background spacetime must be introduced. With a small coupling coefficient, only the left-chiral fermion zero mode is localized on the brane. However, with a large coupling coefficient, the left-chiral fermion zero mode is localized on the brane, and a finite number of resonant massive KK modes of the left- and right-chiral fermions are quasilocalized on the brane. And the number of resonances also increases with the coupling coefficient.
Hence, the massless fermion is localized on the brane consists of just the left-chiral KK mode, and the massive fermions are quasilocalized on the brane consist of the left- and right-chiral fermion KK modes, represented the the 4D Dirac massive fermions. The lifetime of the fermion KK resonant modes decreases with their masses.

\begin{acknowledgments}
This work was supported by the National Natural Science Foundation of China (Grants No. 11305119, No. 11705070, and No. 11405121), the Natural Science Basic Research Plan in Shaanxi Province of China (Program No. 2020JM-198), the Fundamental Research Funds for the Central Universities (Grants No. JB170502), and the 111 Project (B17035).
\end{acknowledgments}


\end{document}